\documentclass[lettersize,journal]{IEEEtran}
\usepackage{amsmath,amsfonts}
\usepackage{algorithmic}
\usepackage{algorithm}
\usepackage{array}
\usepackage{comment}
\usepackage{textcomp}
\usepackage{stfloats}
\usepackage{url}
\usepackage{verbatim}
\usepackage{graphicx}
\usepackage{subcaption}
\usepackage{cite}  % 使用cite包
\usepackage{amsmath}
\usepackage{cuted}
\usepackage{lipsum} 
\usepackage{color}
\usepackage{upgreek}
\usepackage[absolute,overlay]{textpos} % 导入 textpos 包
\newcommand{\triangleq}{\stackrel{\triangle}{=}}
\usepackage[margin=0.5in]{geometry}
\usepackage{caption} % 引入 caption 包来调整字体和大小
\usepackage{times}
\usepackage{titlesec}
\usepackage{multirow}
\titlespacing*{\subsection}{0pt}{*1}{*0.5} % 参数分别是左边距、段前间距、段后间距
\usepackage{indentfirst}
\usepackage{booktabs}
\usepackage[x11names]{xcolor}
\usepackage{comment}
\usepackage{colortbl} % 引入colortbl宏包
\usepackage{xcolor}
\usepackage{xpatch}
\usepackage[T1]{fontenc} % 确保字体编码
\usepackage{textcomp}    % 提供 \textbar
\usepackage{enumitem}
\usepackage{hyperref}     % 再超链接
\usepackage[nameinlink,noabbrev]{cleveref} % 最后 cleveref
\floatname{algorithm}{Workflow}

\makeatletter

\ExplSyntaxOn

\clist_new:N \g_my_blue_bib_clist
\clist_gset:Nn \g_my_blue_bib_clist
  {
    Zeng2019AccessingFromTheSky,
    10955337,
    11072035,
    Mu2023UAVmeetsISAC,
    Jarraya2025GNSSDeniedUAV,
    11153056,
    9898900,
    10600143,
    10962304,
    nlosAdvantage1,
    nlosAdvantage2,
    Zhang2023GeneralChannelISACmag,
    Zhang2024ClusterStatisticalTWCom,
    He2020V2VmmWaveVTM,
    Liu2022JSACISACSurvey,
    Chang2025WCLTHzEnvRecon,
    Shen2024TWCProjective6DPose,
    Du2024IoTJGeneralSLAM,
    Que2023TComJointBeamSLAM,
    RastorguevaFoi2024JSACRadioSLAMe2e,
    Ge2020SensorsClusteringAssignment,
    Kim2024TSPSetTypeBP,
    Yang2023IoTAngleSLAM,
    Yang2022TWcPlugPlayCrowdsourcing,
    Lin2024TWCEnvReconMultiUser,
    Ge2022JSACEKPMBM,
    Yang2023WCMultiDomainCoopSLAM,
    Kim2020TWCCoopPHDMapFusion,
    Yang2022JSACHybridActivePassive,
    Yang2025LWCLISACPrototype,
    Kim2020mmWavePHD,
    Ge2022EKPMBM,
    Du2024mmWaveSLAM,
    Kschischang2001FactorGraph,
    Leitinger2019BP_SLAM,
    Mendrzik2019SituationalAwareness,
    Leitinger2023DataFusionSLAM,
    Gao2024MP_SLAM_Correction,
    Roh2014mmWaveBeamforming,
    Granstrom2023ExtendedObjectTracking,
    Ge2020_5GSLAM_Sensors,
    Kim2022PMBM_VehTech,
    Wielandner2023NonIdealSurfaces,
    Yang2022HybridActivePassiveSLAM,
    Yang2025ISACPrototype,
    Hu2024RadioSLAM,
    mmwaveChanel,
    xie2016overview,
    yang2018channel,
    9693225,
    meyer2018message,
    9585528,
    esposti2007measurement,
    Sun2025RayleighDistance,
    10471226,
    10077117,
    Sun2025LandToShipTWC,
    Zhang2024ISACChannelComMag,
    Liu2024SharedClusterISAC_TVT,
    Zhang2025FourSteps6GAIComMag,
    Pei2025TrackingConvertedRadar,
    11352781
  }

\cs_new_protected:Npn \SetBibColor #1
  {
    \bool_set_false:N \l_tmpa_bool
    \clist_map_inline:Nn \g_my_blue_bib_clist
      {
        \str_if_eq:eeT
          { \tl_to_str:n {#1} }
          { \tl_to_str:n {##1} }
          {
            \bool_set_true:N \l_tmpa_bool
            \clist_map_break:
          }
      }
    \bool_if:NTF \l_tmpa_bool { \color{black} } { \normalcolor }
  }

\ExplSyntaxOff

\let\oldbibitem\bibitem

\def\bibitem{\@ifnextchar[\BibItemWithOpt\BibItemNoOpt}

\def\BibItemNoOpt#1{%
  \SetBibColor{#1}%
  \oldbibitem{#1}%
}

\def\BibItemWithOpt[#1]#2{%
  \SetBibColor{#2}%
  \oldbibitem[#1]{#2}%
}

\makeatother

\usepackage{accents,scalerel}

\usepackage{xcolor}
\definecolor{myPurple}{RGB}{128,0,128} % 典型紫色（#800080）

\usepackage{xcolor}
\usepackage[font=footnotesize]{caption}
\begin{document}
\title{\parbox{\linewidth}{\centering Fusion of Monostatic and Bistatic Sensing for ISAC-Enabled Low-Altitude Environment Mapping}}

\author{{Meihui Liu}, Shu Sun,~\IEEEmembership{Senior Member, IEEE}, {Ruifeng Gao},~\IEEEmembership{{Member, IEEE}}, \\ Jianhua Zhang,~\IEEEmembership{Fellow, IEEE} and Meixia Tao,~\IEEEmembership{Fellow, IEEE}
        % <-this % stops a space
\thanks{M. Liu, S. Sun, and M. Tao are with the School of Information Science and Electronic Engineering, Shanghai Jiao Tong University, Shanghai 200240, China (e-mail: \{meihui\_liu, shusun, mxtao\}@sjtu.edu.cn).}
\thanks{R. Gao is with the School of Transportation and Civil Engineering, Nantong University, Nantong 226019, China (e-mail: grf@ntu.edu.cn).}
\thanks{J. Zhang is with the State Key Laboratory
of Networking and Switching Technology, Beijing University of Posts and
Telecommunications, Beijing 100876, China (e-mail: 
jhzhang@bupt.edu.cn).}
% <-this % stops a space
%\thanks{Manuscript received April 19, 2021; revised August 16, 2021.},
}



\maketitle

\begin{abstract}
Driven by the rapid growth of the low-altitude economy, integrated sensing and communication (ISAC) technologies are essential to meet the stringent demands for reliable connectivity and situational awareness. Within this context, multipath-based simultaneous localization and mapping has emerged as a promising approach by leveraging radio frequency (RF) multipath to reconstruct environment maps alongside agent localization. Nevertheless, existing studies largely confine themselves to bistatic non-line-of-sight links and assume purely specular reflections from smooth surfaces, overlooking the monostatic sensing capability inherent in ISAC systems and the diffuse scattering effects induced by non-ideal outdoor building facades. To address these limitations, this paper presents the first Bayesian multipath-based environment mapping framework for ISAC that integrates monostatic and bistatic measurements under non-ideal surface propagation. We establish geometric relationships linking both sensing modes to a common reflective surface, enabling their association with the same physical feature for data-level fusion. Building on this formulation, we design two complementary Bayesian frameworks with corresponding factor-graph representations, allowing flexible adaptation to different scene requirements. The effectiveness of the proposed approach is validated through synthetic RF data, demonstrating that the fusion of monostatic and bistatic links consistently yields environment maps with higher accuracy, greater robustness and faster convergence than single-link baselines.
\end{abstract}

\begin{IEEEkeywords}
\textcolor{black}{Integrated sensing and communication (ISAC)}, monostatic and bistatic sensing,  multipath-based environment mapping, factor-graph, non-ideal surfaces.
\end{IEEEkeywords}
\vspace{-5pt}
\section{Introduction}
The rapid growth of the low-altitude economy is accelerating the deployment of diverse aerial platforms, including uncrewed aerial vehicles (UAVs) and urban air mobility systems \cite{10955337}. This evolution demands future networks capable of concurrently providing reliable connectivity and precise environmental sensing and mapping \cite{LowAltitudeISACWhitePaper2024,Mu2023UAVmeetsISAC,11072035}. This dual requirement resonates with emerging 6G paradigms that emphasize acquiring wireless environmental information  for environment--channel mapping and proactive air-interface adaptation \cite{Zhang2025FourSteps6GAIComMag}. Integrated sensing and communication (ISAC) addresses this need by unifying connectivity and sensing within shared resources, enabling transmissions to simultaneously convey information and sense the propagation environment \cite{ITU2022FutureIMT2030,Liu2022JSACISACSurvey}.

In building-dense low-altitude corridors, electromagnetic propagation is dominated by non-line-of-sight (NLoS) multipath components. Rather than treating multipath as impairment, ISAC systems can exploit the geometric information embedded within these propagation paths to infer environmental structure, as each multipath component encodes attributes of its associated scatterer, such as position, orientation, and spatial extent \cite{10077117,Leitinger2019BP_SLAM,10962304,Sun2025LandToShipTWC,11352781}. This capability has motivated two methodological paradigms for environment mapping: deterministic geometric reconstruction and probabilistic inference.  Deterministic approaches reconstruct scene geometry by solving the geometric relationships among transmitter, receiver, and scattering objects, constrained by estimated multipath parameters such as delay, angle of departure (AoD), and angle of arrival (AoA).  Recent advances span from 
millimeter-level precision in terahertz (THz) monostatic sensing \cite{11153056},
to sub-decimeter accuracy in THz bistatic fusion systems \cite{Chang2025WCLTHzEnvRecon}, 
and 3D reconstruction via hybrid geometric–learning frameworks 
\cite{Mou2023mmWave3DSLAMarXiv}.  However, the lack of explicit uncertainty quantification in such deterministic approaches limits their robustness in the presence of measurement noise and model mismatch. This limitation is directly addressed by probabilistic frameworks, which cast mapping as a Bayesian inference problem with inherent uncertainty propagation.

The probabilistic framework has been primarily advanced through simultaneous localization and mapping (SLAM), where the core challenge is that the number of map features (MFs, e.g., reflective surfaces, scatterers) is unknown and dynamically evolves, requiring systems to simultaneously infer feature existence, location, and association under uncertainty. Two mainstream paradigms have emerged: random finite set methods model feature collections as ensembles with set priors (e.g., Bernoulli, Poisson), offering theoretical completeness but facing combinatorial explosion in data association \cite{Mahler2014Advances,Kim2020mmWavePHD,Du2024IoTJGeneralSLAM}; message passing methods adopt graphical models with binary existence variables, decomposing global inference into local message exchanges via factor graphs \cite{Leitinger2019BP_SLAM,Kschischang2001FactorGraph}, exhibiting superior scalability for large-scale mapping.

Leveraging these advantages, this paper adopts the message passing framework. Since the seminal work that first performed environment mapping using first-order NLoS multipath delays \cite{Leitinger2019BP_SLAM}, this direction has achieved breakthroughs across multiple dimensions. In terms of observation domain expansion, researchers have incorporated signal amplitude information to enhance multipath resolution under low signal-to-noise ratio conditions \cite{10615625}, and fused multi-source measurements such as AoA and time difference of arrival to construct over-constrained geometric systems \cite{Mendrzik2019SituationalAwareness}. Regarding system architecture, cooperative sensing across base stations (BSs) and distributed data fusion strategies have enabled globally consistent map representations \cite{Leitinger2023DataFusionSLAM,Yang2022JSACHybridActivePassive}. In physical modeling, differentiated treatment of reflecting surfaces and point scatterers has allowed systems to distinguish propagation characteristics of different feature types \cite{10615625}. The above works have primarily focused on first-order specular reflection paths.
Subsequent research has extended the framework to higher-order multipath reflections, exploiting multi-bounce paths via soft ray tracing \cite{Leitinger2023DataFusionSLAM} or geometric models \cite{Hu2024RadioSLAM}.

Despite these advances, a key limitation remains prevalent in many existing formulations: exclusive association constraints. Existing methods commonly assume that each MF generates at most one multipath component (MPC) and each MPC originates from at most one MF.  This one-to-one data association (DA) model is rooted in the ideal specular-reflection assumption, yet it fails to capture non-ideal scattering, where a single MF (e.g., a rough building facade) can give rise to multiple resolvable MPCs.  Modern wideband systems with massive arrays \cite{11153056,Sun2025RayleighDistance,9898900}, further expose this issue by resolving not only dominant specular paths but also weaker non-specular components, thereby violating the exclusivity assumption and causing biased estimates and spurious features under traditional DA models.

To mitigate this one-to-many DA challenge, several solutions have emerged. In \cite{Ge2020_5GSLAM_Sensors}, clustering methods group measurements to simplify association. Building on this, \cite{Kim2022PMBM_VehTech} introduced unified models incorporating both surfaces and scatterers with type indicators. Related non-ideal surface treatments have also been incorporated into message-passing formulations: \cite{Wielandner2023NonIdealSurfaces} modeled non-ideal surface effects in delay and amplitude domains, later extended to multiple-input multiple-output (MIMO) scenarios \cite{Wielandner2024MIMO_NonIdealSurfaces} to leverage angular information. More recently, \cite{10962304} addressed compound scenarios with coexisting non-ideal surfaces and large-sized reflectors. While these methods substantially alleviate DA ambiguity, they are largely confined to bistatic sensing. In ISAC, sensing may also be realized in a monostatic configuration with co-located transmit and receive chains \cite{Zhang2024ISACChannelComMag,Liu2024SharedClusterISAC_TVT}.

Notably, the same high-resolution massive MIMO systems that densify multipath also make monostatic backscatter sufficiently resolvable for environmental geometry perception. Emerging ISAC BSs, equipped with advanced RF front-ends and large arrays, can reliably extract strong backscatter components from reflective surfaces, which encode key geometric attributes such as surface position and orientation. 
Moreover, monostatic observation offers two inherent advantages: geometric stability due to the fixed BS-centric sensing geometry, and measurement superiority stemming from BS-grade hardware.  By jointly fusing heterogeneous monostatic and bistatic observations, the proposed framework improves reconstruction accuracy and robustness over bistatic-only sensing by injecting stable surface pose constraints from monostatic backscatter into bistatic inference.

However, monostatic observations remain largely unexploited in existing multipath-based environment mapping, and their data-level fusion with bistatic links is virtually absent. Although \cite{Yang2022JSACHybridActivePassive} proposed a hybrid architecture, it employs monostatic links solely for feature-level initialization rather than continuous data-level fusion, and relies on exclusive association assumptions incompatible with non-ideal surface scattering.  The fundamental gap therefore remains: developing a unified framework that jointly exploits monostatic surface characterization and bistatic geometric constraints while handling one-to-many associations from non-ideal scattering.

This paper addresses this challenge by establishing a Bayesian multipath-based environment mapping framework that integrates monostatic and bistatic sensing under non-ideal surface conditions. Our key contributions are:
\begin{itemize}[label=$\bullet$]
  \item  We establish a unified geometric association that bridges monostatic and bistatic observations through a common MF, enabling joint constraint of the same physical facade from heterogeneous link geometries. On this basis, we construct  the corresponding observation likelihoods within a consistent Bayesian framework to achieve coherent data-level fusion.
  \item We develop two Bayesian fusion schemes that incorporate high-quality monostatic observations absent in existing approaches. Both handle one-to-many DA problem under non-ideal scattering with comparable accuracy. Scheme I enables parallel processing for low latency, while Scheme II employs cross-link sequential processing for enhanced completeness, collectively addressing diverse deployment needs.
  \item We provide factor graph representations for both schemes and conduct systematic  complexity analysis for sum-product inference. Through extensive simulations using synthetic RF data generated by widely-employed ray-tracing software, we demonstrate consistent improvements in mapping accuracy over existing bistatic-only and feature-level fusion baselines, validating the benefits of coherent data-level monostatic-bistatic integration.
\end{itemize}

The rest of the paper is organized as follows. Section II presents the
system model and statistical formulations. Section III develops the joint posterior probability density function (PDF) and factor graph for both fusion schemes. Section IV presents the inference algorithm and analyzes its computational complexity. Section V demonstrates performance advantages through numerical results, and Section VI concludes the paper.

\textit{Notation:} Random variables and random vectors are explicitly identified at their first appearance; unless otherwise specified, symbols denote deterministic quantities. Deterministic scalars are written in italic font, while deterministic vectors and matrices are denoted by bold lowercase and uppercase letters, respectively.  Furthermore, \(\boldsymbol{x}^{\mathrm{T}}\) denotes the transpose of vector \(\boldsymbol{x}\); \(\propto\) indicates equality up to a normalization factor. \(f(\boldsymbol{x})\) denotes the PDF of random vector \(\boldsymbol{x}\); \(f(\boldsymbol{x} \mid \boldsymbol{y})\) denotes the conditional PDF of random vector \(\boldsymbol{x}\) conditioned on random vector \(\boldsymbol{y}\).  \(\mathcal{N}(\boldsymbol{x}; \boldsymbol{\mu}, \boldsymbol{\Sigma})\) denotes the Gaussian PDF (of random vector \(\boldsymbol{x}\)) with mean \(\boldsymbol{\mu}\) and covariance matrix \(\boldsymbol{\Sigma}\). \textcolor{black}{$\mathcal{U}(x; a,b)$ denotes the uniform PDF of a scalar $x$ on the interval $[a,b]$.} 

\section{System Model}
\begin{figure}
    \centering
    \includegraphics[width=1\linewidth]{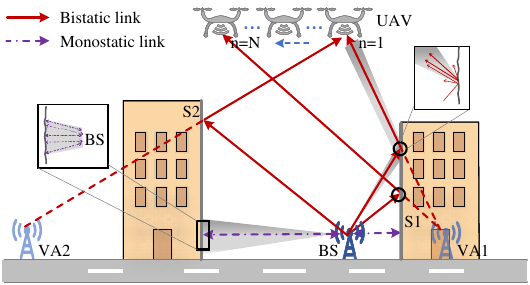}
    \caption{Illustration of the environment mapping scenario with monostatic and bistatic links. VA1 and VA2 are the mirror-symmetric points of the BS about the non-ideal building facades S1 and S2, respectively, serving as their corresponding mapping features MF1 and MF2.}
    \label{fig:placeholder}
    \vspace{-5pt}
\end{figure}

In this section, we introduce the scenario and describe in detail how to associate data from both bistatic and monostatic links to the same MF. An illustrative example is depicted in Fig. \ref{fig:placeholder}.

\subsection{Scenario Description}
We consider a low-altitude ISAC downlink scenario in which a UAV is served by $J$ ground BSs, with the known position of BS~$j$ denoted by $\mathbf{p}_{\mathrm{BS}}^{(j)}$, $j=1,\ldots,J$. The UAV trajectory is assumed to be obtained from onboard sensors such as an inertial measurement unit and a global positioning system receiver.  Therefore, we focus on environment mapping rather than simultaneous localization and mapping. \textcolor{black}{At sensing epoch 
$n$, the UAV position is denoted by 
$\mathbf{u}_{n}\in\mathbb{R}^{3}$
, which is assumed known in this work.} Each BS employs an ISAC transceiver enabling monostatic sensing, while the UAV operates in receive mode for data decoding and bistatic sensing; hence, both monostatic (BS-side) and bistatic (UAV-side) observations are available. \textcolor{black}{Each BS is equipped with a calibrated planar array (e.g., a UPA) to enable joint azimuth/elevation parameter estimation; the proposed framework uses only the resulting MPC parameter estimates and their uncertainty.}

The sensing objective is to reconstruct an environment map by estimating the geometry of static building facades. We use virtual anchors (VAs), defined as mirror images of BSs with respect to facades~\cite{Leitinger2019BP_SLAM}, \cite{Mendrzik2019SituationalAwareness}, as compact descriptors. The VA locations are estimated by fusing monostatic and bistatic multipath observations (Section~IV), which in turn enables inference of facade positions; the resulting map can further support communication optimization (e.g., predictive beamforming).

\textcolor{black}{\subsection{ISAC MPC Acquisition and Measurement Models}}
\textcolor{black}{
We consider a communication-centric ISAC mode where the required MPC parameters  are extracted by reusing standard downlink reference/pilot resources (e.g., SSB-like signaling and CSI/DMRS/DL-PRS-like pilots), transmitted via wide beams or beam sweeping for coverage, rather than introducing a dedicated radar-only waveform.
When monostatic sensing is needed, the BS allocates a short receive/listening window to collect backscattered signals, enabled by full-duplex operation or TDD switching.
More general dual-functional ISAC realizations with flexible sensing/communication resource allocation across time, frequency, or space are also applicable; they mainly affect the achievable estimation accuracy while producing the same estimator outputs required by our Bayesian mapping and fusion.
We assume that bistatic and monostatic MPC parameter sets are extracted synchronously within each sensing epoch: during an ISAC downlink burst, the UAV estimates bistatic delays, whereas the BS estimates monostatic delay and AoA parameters via array processing.
The sensing--communication coupling is reflected by the pilot/beam-training density and the listening time, which jointly determine the estimation uncertainty (and the corresponding resource overhead).}

\textcolor{black}{Regardless of the specific ISAC realization, the received baseband signals on both links admit the same parametric multipath decomposition: a set of dominant specular MPCs, each of which may be accompanied by a local cluster of weaker diffuse subpaths arising from non-ideal reflective surfaces, as supported by measurement-based and model-based studies~\cite{10962304,Ge2020_5GSLAM_Sensors,Wielandner2023NonIdealSurfaces,Wielandner2024MIMO_NonIdealSurfaces,esposti2007measurement,ZhangDiffuseScatteringNPJ_toappear}. }
This motivates the unified multipath model:
\begin{equation}
\mathbf{s}_{\mathrm{rx}}(t)
=
\sum_{l=1}^{L}
\alpha_{l}
\!\left(
    \mathbf{x}(t-\tau_l)
    +
    \sum_{i=1}^{S_{l}}
        \beta_{l,i}\,
        \mathbf{x}(t-\tau_l-\nu_{l,i})
\right)
+
\mathbf{n}(t),
\label{eq:clustered}
\end{equation}
\textcolor{black}{
where $l$ indexes the dominant specular components and $S_l$ denotes the number of diffuse subpaths associated with component $l$.
In the monostatic case, each specular component is interpreted as a dominant return that can be explained by reflection from a candidate facade plane.
In the bistatic case, the specular set includes facade reflections and may also include a LoS component, when present; a LoS component is modeled as a pure specular path and does not generate a diffuse cluster, i.e., $S_l=0$ for the LoS path.
The coefficient $\alpha_l$ denotes the complex gain of the $l$th specular component, while $\beta_{l,i}$ is the relative complex attenuation of the $i$th diffuse subpath within cluster $l$.
The delay $\tau_l$ is the specular-path delay, and $\nu_{l,i}>0$ is the excess delay of the $i$th diffuse subpath, i.e., $\tau_{l,i}=\tau_l+\nu_{l,i}$.
The received baseband signal $\mathbf{s}_{\mathrm{rx}}(t)\in\mathbb{C}^{N_{\mathrm r}}$ is an \emph{array snapshot} that stacks the complex samples across the $N_{\mathrm r}$ receive antennas at time $t$. 
The term $\mathbf{x}(t-\tau)\in\mathbb{C}^{N_{\mathrm r}}$ denotes the \emph{per-antenna} contribution of a single MPC with delay $\tau$ to this array snapshot, i.e., the rank-one array response induced by that path prior to any receive combining (beamforming).\footnote{\textcolor{black}{For completeness, a common narrowband array-response form with planar arrays is
$\mathbf{x}(t-\tau)=\mathbf{a}_{\mathrm r}(\mathbf{\Theta}^{\mathrm r})\mathbf{a}_{\mathrm t}^{\mathrm H}(\mathbf{\Theta}^{\mathrm t})\mathbf{s}_{\mathrm{tx}}(t-\tau)$,
where $\mathbf{s}_{\mathrm{tx}}(t)\in\mathbb{C}^{N_{\mathrm t}}$ is the beamformed transmit signal across $N_{\mathrm t}$ antennas,
$\mathbf{a}_{\mathrm r}(\cdot)\in\mathbb{C}^{N_{\mathrm r}}$ and $\mathbf{a}_{\mathrm t}(\cdot)\in\mathbb{C}^{N_{\mathrm t}}$ are the receive/transmit steering vectors, and
$\mathbf{\Theta}^{\mathrm r}=[\theta^{\mathrm r},\phi^{\mathrm r}]^{\mathsf T}$ (resp.\ $\mathbf{\Theta}^{\mathrm t}$) denotes the azimuth--elevation pair of the AoA (resp.\ AoD).}}
This signal model is introduced only to motivate the clustered MPC structure; in the sequel, the proposed mapping and fusion use only the detected MPC parameter estimates (e.g., delays/angles) and their associated uncertainties as measurements, without explicitly relying on the waveform-/steering-vector-level expansion above.
The term $\mathbf{n}(t)\in\mathbb{C}^{N_{\mathrm r}}$ denotes additive receiver noise.
At each sensing epoch \(n\), the UAV and each BS apply a channel-parameter estimator~\cite{xie2016overview,yang2018channel,9693225} to \(\mathbf{s}_{\mathrm{rx}}(t)\), yielding detected MPC parameter estimates and associated uncertainties that define the bistatic and monostatic measurement models in the sequel.} \textcolor{black}{In the proposed framework, we retain only MPCs that admit a single-bounce interpretation. Multi-bounce components are not explicitly modeled and are either suppressed during screening or absorbed into clutter, outliers, or model mismatch~\cite{10600143}.
} 

% ---------- Bistatic pseudo-range measurements ----------
With a single-antenna UAV, \textcolor{black}{for each detected bistatic MPC (indexed by $m$ at time $n$) associated with BS~$j$, the channel-parameter estimator provides an observed delay estimate ${\hat \tau}^{\mathrm{bi},(j)}_{m,n}$. We regard ${\hat \tau}^{\mathrm{bi},(j)}_{m,n}$ as one realization of the estimator-output delay random variable $T^{\mathrm{bi},(j)}_{m,n}$. Conditioned on the cluster association $G(m)$, we model
\begin{equation}
T^{\mathrm{bi},(j)}_{m,n}
=
{\tau}^{\mathrm{bi},(j)}_{G(m),n}
+
V^{(j)}_{m,n}
+
E^{\mathrm{bi},(j)}_{m,n},
\label{eq:delayModelBi}
\end{equation}
where ${\tau}^{\mathrm{bi},(j)}_{G(m),n}$ is the true (unknown) specular-path delay of the associated cluster. The random variable $V^{(j)}_{m,n}$ captures the physical excess delay of a diffuse subpath relative to the specular component: $V^{(j)}_{m,n}=0$ for specular components and $V^{(j)}_{m,n}>0$ for diffuse subpaths. The estimator-induced delay error is modeled as $E^{\mathrm{bi},(j)}_{m,n}\sim\mathcal N(0,\sigma_\tau^2)$.
We then define the scalar pseudo-range random variable and its observed realization as
\begin{equation}
Z^{\mathrm{bi},(j)}_{m,n}\triangleq \mathrm c\,T^{\mathrm{bi},(j)}_{m,n},
\qquad
z^{\mathrm{bi},(j)}_{m,n}\triangleq \mathrm c\,\hat{\tau}^{\mathrm{bi},(j)}_{m,n}.
\label{eq:pseudorangeDefBi}
\end{equation}
where $\mathrm{c}$ denotes the speed of light. Therefore, conditioned on $G(m)$ and $V^{(j)}_{m,n}$, the pseudo-range is conditionally Gaussian:
\begin{equation}
Z^{\mathrm{bi},(j)}_{m,n}\,\big|\,G(m),V^{(j)}_{m,n}
\sim
\mathcal N\!\Big(
\mathrm c\big({\tau}^{\mathrm{bi},(j)}_{G(m),n}+V^{(j)}_{m,n}\big),\;
\mathrm c^2\sigma_\tau^2
\Big).
\label{eq:ZcondBiGaussian}
\end{equation}
The~\eqref{eq:ZcondBiGaussian} makes the estimator-error propagation explicit: the delay-domain uncertainty $\sigma_\tau^2$ is mapped to the pseudo-range domain as $\sigma_z^2=\mathrm c^2\sigma_\tau^2$. In Sec.~II-C, the pseudo-range measurement is linked to the environmental facades through the bistatic geometry, leading to the corresponding likelihood function in Sec.~III-C; the propagated uncertainty then enters the proposed Bayesian inference framework via this likelihood.}
\noindent\textcolor{black}{It is worth noting that any residual BS--UAV timing offset on link $j$ is absorbed into $E^{\mathrm{bi},(j)}_{m,n}$ (equivalently, into $\sigma_\tau^2$). If present, a slowly varying offset can also be modeled as a per-BS nuisance parameter and inferred within the same factor-graph framework.}

% ---------- Monostatic pseudo-position measurements ----------
\textcolor{black}{With a multi-antenna BS, for each detected monostatic MPC (indexed by $m$ at time $n$) associated with BS~$j$, the channel-parameter estimator produces an observed delay/angle estimate vector
$
\hat{\boldsymbol{\eta}}^{\mathrm{mo},(j)}_{m,n}
\triangleq
\big[\hat{\tau}^{\mathrm{mo},(j)}_{m,n},\,\hat{\theta}^{\mathrm{mo},(j)}_{m,n},\,\hat{\phi}^{\mathrm{mo},(j)}_{m,n}\big]^{\mathrm T}.
$
Let
$
\boldsymbol{\eta}^{\mathrm{mo},(j)}_{m,n}
\triangleq
\big[\tau^{\mathrm{mo},(j)}_{m,n},\,\theta^{\mathrm{mo},(j)}_{m,n},\,\phi^{\mathrm{mo},(j)}_{m,n}\big]^{\mathrm T}
$
denote the corresponding true but unknown delay/angle triplet. We regard $\hat{\boldsymbol{\eta}}^{\mathrm{mo},(j)}_{m,n}$ as one realization of the estimator-output random vector $\boldsymbol{T}^{\mathrm{mo},(j)}_{m,n}\in\mathbb{R}^3$ and model
\begin{equation}
\boldsymbol{T}^{\mathrm{mo},(j)}_{m,n}
=
\boldsymbol{\eta}^{\mathrm{mo},(j)}_{m,n}
+
\boldsymbol{E}^{\eta,\mathrm{mo},(j)}_{m,n},
\label{eq:etaModelMo}
\end{equation}
where $\boldsymbol{E}^{\eta,\mathrm{mo},(j)}_{m,n}$ denotes the front-end delay/angle estimation error. When available, we approximate it as conditionally Gaussian,
$
\boldsymbol{E}^{\eta,\mathrm{mo},(j)}_{m,n}\,|\,\boldsymbol{\eta}^{\mathrm{mo},(j)}_{m,n}
\sim
\mathcal N\!\left(\mathbf 0,\boldsymbol{\Sigma}^{(j)}_{\eta,m,n}\right),
$
with covariance $\boldsymbol{\Sigma}^{(j)}_{\eta,m,n}$.
Define the delay/angle-to-position conversion mapping
\begin{equation}
\mathbf g^{(j)}(\tau,\theta,\phi)
\triangleq
\mathbf{p}^{(j)}_{\mathrm{BS}}
+\frac{\mathrm c}{2}\tau\,\mathbf u(\theta,\phi),
\label{eq:gDefMo}
\end{equation}
where
$
\mathbf{u}(\theta,\phi)=
\big[\sin\phi\cos\theta,\ \sin\phi\sin\theta,\ \cos\phi\big]^{\mathrm T}.
$
Accordingly, the monostatic pseudo-position random vector and its observed realization can be written as
\begin{equation}
\boldsymbol{Z}^{\mathrm{mo},(j)}_{m,n}
\triangleq
\mathbf g^{(j)}\!\left(\boldsymbol{T}^{\mathrm{mo},(j)}_{m,n}\right),
\qquad
\boldsymbol{z}^{\mathrm{mo},(j)}_{m,n}
\triangleq
\mathbf g^{(j)}\!\left(\hat{\boldsymbol{\eta}}^{\mathrm{mo},(j)}_{m,n}\right),
\label{eq:pseudoposDefMo}
\end{equation}
i.e.,
$
\boldsymbol{z}^{\mathrm{mo},(j)}_{m,n}
=
\mathbf{p}^{(j)}_{\mathrm{BS}}
+\frac{\mathrm c}{2}\,\hat{\tau}^{\mathrm{mo},(j)}_{m,n}\,
\mathbf{u}\!\left(\hat{\theta}^{\mathrm{mo},(j)}_{m,n},\hat{\phi}^{\mathrm{mo},(j)}_{m,n}\right).
$
Since $\mathbf g^{(j)}(\cdot)$ is nonlinear, the induced pseudo-position error is generally non-additive and non-Gaussian. To obtain a tractable likelihood, we adopt a first-order converted-measurement approximation \cite{Pei2025TrackingConvertedRadar,Yang2022JSACHybridActivePassive}: linearizing $\mathbf g^{(j)}(\cdot)$ around a local operating point yields
\[
\boldsymbol{Z}^{\mathrm{mo},(j)}_{m,n}
\approx
\mathbf g^{(j)}\!\left(\boldsymbol{\eta}^{\mathrm{mo},(j)}_{m,n}\right)
+
\mathbf J^{(j)}_{g,m,n}\,
\boldsymbol{E}^{\eta,\mathrm{mo},(j)}_{m,n},
\]
where $\mathbf J^{(j)}_{g,m,n}$ is the Jacobian of $\mathbf g^{(j)}$ at the chosen linearization point. This motivates the following Gaussian-equivalent pseudo-position-domain measurement model:
\begin{equation}
\boldsymbol{Z}^{\mathrm{mo},(j)}_{m,n}
\approx
\mathbf{p}^{(j)}_{\mathrm{BS}}
+\frac{\mathrm c}{2}\,{\tau}^{\mathrm{mo},(j)}_{m,n}\,
\mathbf{u}\!\left({\theta}^{\mathrm{mo},(j)}_{m,n},{\phi}^{\mathrm{mo},(j)}_{m,n}\right)
+
\boldsymbol{E}^{\mathrm{mo},(j)}_{m,n},
\label{eq:measureModelMo}
\end{equation}
with
$
\boldsymbol{E}^{\mathrm{mo},(j)}_{m,n}
\approx
\mathbf J^{(j)}_{g,m,n}\boldsymbol{E}^{\eta,\mathrm{mo},(j)}_{m,n}
$
and covariance
\[
\mathbf R^{(j)}_{m,n}
\triangleq
\mathrm{Cov}\!\left(\boldsymbol{E}^{\mathrm{mo},(j)}_{m,n}\right)
\approx
\mathbf J^{(j)}_{g,m,n}\,
\boldsymbol{\Sigma}^{(j)}_{\eta,m,n}\,
\big(\mathbf J^{(j)}_{g,m,n}\big)^{\mathrm T},
\]
via standard first-order uncertainty propagation \cite{Pei2025TrackingConvertedRadar}. Accordingly, we model $\boldsymbol{E}^{\mathrm{mo},(j)}_{m,n}$ in a Gaussian-equivalent (moment-matched) sense as
$
\boldsymbol{E}^{\mathrm{mo},(j)}_{m,n}\sim\mathcal N(\mathbf 0,\mathbf R^{(j)}_{m,n}).
$
Even when the true errors deviate from Gaussianity, the propagation from delay/angle estimates to pseudo-position and then to the geometric residual follows the same deterministic mappings. The covariance $\mathbf R^{(j)}_{m,n}$ is further propagated to the monostatic geometric likelihood through the projection-residual covariance (defined later in Sec.~II-C), so larger converted-measurement uncertainty yields a broader likelihood term and automatically down-weights unreliable monostatic pseudo-positions in the Bayesian fusion. If needed, $\mathbf R^{(j)}_{m,n}$ can be conservatively inflated (e.g., based on calibration or prior knowledge) to improve robustness to residual bias and non-Gaussian estimation artifacts.}

\textcolor{black}{
For BS~$j$ at sensing epoch $n$, let $M^{(j)}_{n,\mathrm{bi}}$ and $M^{(j)}_{n,\mathrm{mo}}$ denote the numbers of detected MPCs on the bistatic and monostatic links, respectively. The corresponding stacked pseudo-measurements are defined as
$
\boldsymbol{z}^{(j)}_{n}
\triangleq
\big[z^{\mathrm{bi},(j)}_{1,n},\,\ldots,\,z^{\mathrm{bi},(j)}_{M^{(j)}_{n,\mathrm{bi}},n}\big]^{\mathrm T}
\in \mathbb{R}^{M^{(j)}_{n,\mathrm{bi}}},
$
and
$
\tilde{\boldsymbol{z}}^{(j)}_{n}
\triangleq
\big[(\boldsymbol{z}^{\mathrm{mo},(j)}_{1,n})^{\mathrm T},\,\ldots,\,(\boldsymbol{z}^{\mathrm{mo},(j)}_{M^{(j)}_{n,\mathrm{mo}},n})^{\mathrm T}\big]^{\mathrm T}
\in \mathbb{R}^{3M^{(j)}_{n,\mathrm{mo}}},
$
where \(\boldsymbol{z}^{\mathrm{mo},(j)}_{m,n}\in\mathbb{R}^{3}\). When no ambiguity arises, we write \(z^{(j)}_{m,n}\triangleq z^{\mathrm{bi},(j)}_{m,n}\) and \(\tilde{\boldsymbol{z}}^{(j)}_{m,n}\triangleq \boldsymbol{z}^{\mathrm{mo},(j)}_{m,n}\).
}

\begin{table*}[t]
{\color{black}
\caption{\textcolor{black}{Notations of important variables.}}
\label{tab:notation_compact}
\centering
\footnotesize
\setlength{\tabcolsep}{4pt}
\renewcommand{\arraystretch}{0.92}
\begin{tabular}{@{}p{0.125\textwidth} p{0.35\textwidth} p{0.125\textwidth} p{0.35\textwidth}@{}}
\hline
\textbf{Notation} & \textbf{Definition} & \textbf{Notation} & \textbf{Definition} \\
\hline

$\mathbf p_{\mathrm{BS}}^{(j)}$ & Known position of BS $j$.
        & $\mathbf u_n$ & UAV position at epoch $n$ (assumed known). \\

$\sigma_\tau^2$ & Variance of bistatic delay error.
        & $\sigma_z^2$ & Pseudo-range variance, $\sigma_z^2=\mathrm c^2\sigma_\tau^2$ \\

$\mathbf R^{(j)}_{m,n}$ & Converted-measurement covariance of pseudo-position.
        & $\boldsymbol{\Sigma}^{(j)}_{\mathbf e,m,n,k}(\cdot)$ & Residual covariance induced by $\mathbf R^{(j)}_{m,n}$. \\

$\Pi^{(j)}_{k,n}$ & VA-induced reflector plane for hypothesis $k$.
        & $\boldsymbol e^{(j)}_{m,n,k}$ & Projection residual. \\

$\boldsymbol x^{(j)}_{k,n}$ / $\tilde{\boldsymbol x}^{(j)}_{k,n}$ & Latent bistatic / monostatic PMF  position state.
        & $r^{(j)}_{k,n}$ /  $\tilde{r}^{(j)}_{k,n}$ & Existence indicator of bistatic / monostatic PMF $(j,k)$. \\

$\boldsymbol b_n^{(j)}$ & Bistatic association vector.
        & $\boldsymbol c_n^{(j)}$ & Monostatic association vector. \\

$\boldsymbol y^{(j)}_{k,n}$ / $\tilde{\boldsymbol{y}}^{(j)}_{k,n}$ & Augmented bistatic/monostatic PMF state.
        & $\underline{(\cdot)}$ / $\overline{(\cdot)}$ & Legacy/New-PMF notation. \\

$\mathcal K^{(j)}_{n,\mathrm{bi}}$ / $\mathcal K^{(j)}_{n,\mathrm{mo}}$ & Bistatic/monostatic PMF index set at epoch $n$.
        & $\mathcal M^{(j)}_{n,\mathrm{bi}}$ / $\mathcal M^{(j)}_{n,\mathrm{mo}}$ & Bistatic/monostatic measurement set at epoch $n$ . \\

\hline
\end{tabular}
}
\vspace{-2mm}
\end{table*}
\subsection{Geometric Model}
Having established in \eqref{eq:pseudorangeDefBi} and \eqref{eq:measureModelMo} the relationships between the noisy pseudo-range/pseudo-position observations and the underlying unknown MPC parameters (e.g., delays and angles), we next map these MPC parameters to the environmental geometry through a set of latent VAs. Under this mapping, each pseudo measurement contributes to the Bayesian likelihood with an explicit uncertainty characterization, as formalized in Sec.~III-C. If bistatic and monostatic pseudo measurements arise from the same physical facade, they share the same VA geometry and thus provide mutually consistent constraints on the underlying reflector. Their unified Bayesian fusion then enables direct inference of the VA geometry from noisy pseudo-range and pseudo-position observations.
\textcolor{black}{
For BS~$j$, let $\mathbf p_{\mathrm{VA},k,n}^{(j)}\in\mathbb R^3$ denote the VA location associated with the $k$th facade at time $n$, which serves as the geometric parameter in the VA-based mappings below. In Sec.~III, the corresponding bistatic and monostatic latent state vectors are denoted by $\boldsymbol{x}_{k,n}^{(j)}$ and $\tilde{\boldsymbol{x}}_{k,n}^{(j)}$, respectively; both represent the same underlying VA geometry under the two sensing views.}

\textcolor{black}{
For the bistatic link, if the $m$th detected MPC at time $n$ is associated with facade $k$, then the specular geometric pseudo-range induced by the VA location $\mathbf p_{\mathrm{VA},k,n}^{(j)}$ is
$
\mu_{k,n}^{(j)}
\triangleq \mathrm c\,{\tau}^{\mathrm{bi},(j)}_{G(m),n}
=
\big\|\mathbf u_n-\mathbf p_{\mathrm{VA},k,n}^{(j)}\big\|.
$
Using the pseudo-range measurement model in Sec.~II-B, the bistatic geometric model can be written as
\begin{equation}
Z_{m,n}^{(j)}
=
\mu_{k,n}^{(j)}
+
V_{z,m,n}^{(j)}
+
E_{z,m,n}^{\mathrm{bi},(j)},
\label{eq:bi_geo_meas_relation}
\end{equation}
where $V_{z,m,n}^{(j)}=\mathrm cV_{m,n}^{(j)}$ and $E_{z,m,n}^{\mathrm{bi},(j)}=\mathrm cE_{m,n}^{\mathrm{bi},(j)}$, with $\sigma_z^2=\mathrm c^2\sigma_\tau^2$ as in Sec.~II-B, which enters the bistatic likelihood in Sec.~III-C.
}
\vspace{-2pt}

\textcolor{black}{
For the monostatic link, each detected MPC is mapped to an effective scattering point, whose pseudo-position observation is denoted by $\tilde{\boldsymbol z}_{m,n}^{(j)}$ and is constructed from the estimated delay/angle parameters as in \eqref{eq:measureModelMo}. Under the planar-facade assumption, the noise-free scattering point lies on the reflector plane associated with the facade.
To exploit this structure, for a hypothesized VA location $\mathbf p_{\mathrm{VA},k,n}^{(j)}$,
we define the VA-induced reflector plane
$
\Pi^{(j)}_{k,n}\ \triangleq\ \Pi^{(j)}\!\big(\mathbf p_{\mathrm{VA},k,n}^{(j)}\big)
$
as the perpendicular-bisector plane of the segment connecting the known BS position
$\mathbf p_{\mathrm{BS}}^{(j)}$ and $\mathbf p_{\mathrm{VA},k,n}^{(j)}$.
The corresponding unit normal vector and a point on the plane are defined as
$
\mathbf n^{(j)}_{k,n}\ 
=
\frac{\mathbf p_{\mathrm{BS}}^{(j)}-\mathbf p_{\mathrm{VA},k,n}^{(j)}}{\big\|\mathbf p_{\mathrm{BS}}^{(j)}-\mathbf p_{\mathrm{VA},k,n}^{(j)}\big\|},
\,
\mathbf p^{(j)}_{\mathrm{c},k,n}\ 
=
\frac{1}{2}\Big(\mathbf p_{\mathrm{BS}}^{(j)}+\mathbf p_{\mathrm{VA},k,n}^{(j)}\Big).
$
Accordingly, the plane can be represented as
$
\Pi^{(j)}_{k,n}
=
\Big\{\mathbf r\in\mathbb R^3:\;
\big(\mathbf n^{(j)}_{k,n}\big)^{\mathrm T}
\big(\mathbf r-\mathbf p^{(j)}_{\mathrm{c},k,n}\big)=0
\Big\}.
$
For any point $\mathbf x\in\mathbb R^{3}$, the orthogonal projection onto $\Pi^{(j)}_{k,n}$ is
\begin{equation}
\mathrm{Proj}_{\Pi^{(j)}_{k,n}}(\mathbf x)
\triangleq
\mathbf x
-
\mathbf n^{(j)}_{k,n}\,
\big(\mathbf n^{(j)}_{k,n}\big)^{\mathrm T}
\Big(\mathbf x-\mathbf p^{(j)}_{\mathrm{c},k,n}\Big).
\label{eq:plane_projection}
\end{equation}
In the noise-free case, the monostatic pseudo-position lies on the reflector plane, i.e.,
$
\tilde{\boldsymbol z}_{m,n}^{(j)}
=
\mathrm{Proj}_{\Pi^{(j)}_{k,n}}
\!\Big(\tilde{\boldsymbol z}_{m,n}^{(j)}\Big).
$
With measurement/conversion errors, for the hypothesized VA indexed by $k$, we characterize the deviation from $\Pi^{(j)}_{k,n}$ by the projection residual
\begin{equation}
{\boldsymbol  {e}}^{(j)}_{m,n,k}
\triangleq
\tilde{\boldsymbol z}_{m,n}^{(j)}
-
\mathrm{Proj}_{\Pi^{(j)}_{k,n}}
\!\Big(\tilde{\boldsymbol z}_{m,n}^{(j)}\Big).
\label{eq:mono_proj_residual}
\end{equation}
This residual provides the monostatic geometric consistency measure for the hypothesized VA location $\mathbf p_{\mathrm{VA},k,n}^{(j)}$. Substituting \eqref{eq:measureModelMo} into \eqref{eq:mono_proj_residual} and using \eqref{eq:plane_projection}, we define the residual random vector
$
\boldsymbol{E}^{(j)}_{m,n,k}
\triangleq
\boldsymbol{Z}^{\mathrm{mo},(j)}_{m,n}
-
\mathrm{Proj}_{\Pi^{(j)}_{k,n}}\!\Big(\boldsymbol{Z}^{\mathrm{mo},(j)}_{m,n}\Big).
$
Under the correct plane hypothesis, $\boldsymbol{E}^{(j)}_{m,n,k}$ is induced purely by the pseudo-position error and satisfies
$
\boldsymbol{E}^{(j)}_{m,n,k}
=
\mathbf P^{(j)}_{k,n}\boldsymbol{E}^{\mathrm{mo},(j)}_{m,n},
\,
\mathbf P^{(j)}_{k,n}\triangleq
\mathbf n^{(j)}_{k,n}\big(\mathbf n^{(j)}_{k,n}\big)^{\mathrm T},
$
with $\boldsymbol{E}^{\mathrm{mo},(j)}_{m,n}\sim\mathcal N(\mathbf 0,\mathbf R^{(j)}_{m,n})$. Therefore,
$
\boldsymbol{E}^{(j)}_{m,n,k}\sim
\mathcal N\!\left(\mathbf 0,\ \boldsymbol{\Sigma}^{(j)}_{e,m,n,k}\right),
\,
\boldsymbol{\Sigma}^{(j)}_{e,m,n,k}
=
\mathbf P^{(j)}_{k,n}\,\mathbf R^{(j)}_{m,n}\,\mathbf P^{(j)}_{k,n}.
$
The covariance $\boldsymbol{\Sigma}^{(j)}_{e,m,n,k}$, induced by $\mathbf R^{(j)}_{m,n}$, is used to construct the monostatic geometric likelihood in Sec.~III-C.}

\section{Statistical Model}
Building on the geometric relationships between bistatic and monostatic link measurements and the common VAs established in Sec.~II, we now formulate a unified statistical model to enable subsequent probabilistic inference.

\textcolor{black}{\noindent\textbf{Notation:}
Throughout Sec.~III, lowercase (bold) symbols (e.g., $\boldsymbol{x}^{(j)}_{k,n}$, $r^{(j)}_{k,n}$, $\boldsymbol{y}^{(j)}_{k,n}$) denote latent state random variables/vectors to be inferred. For notational brevity, the same symbols are also used to represent generic values of these variables when they appear as arguments in PDFs/likelihoods (a standard abuse of notation). Accordingly, we write $f(z\mid x)$ as a shorthand for $f_{Z\mid x}(z\mid x)$ evaluated at the observed measurement $z$.} 

\subsection{State Space Model}
We model the VA of BS~$j$ with respect to a building facade as an MF. At time $n$, potential map features (PMFs) observable via the bistatic (BS-to-UAV) link are indexed by $(j,k)$, where $j\in\{1,\ldots,J\}$ denotes the BS index and $k\in\mathcal{K}^{(j)}_{n,\mathrm{bi}}\triangleq\{1,\ldots,K^{(j)}_{n,\mathrm{bi}}\}$ indexes the PMFs associated with BS~$j$; hence, BS~$j$ has $K^{(j)}_{n,\mathrm{bi}}$ bistatic PMFs at time $n$. The existence of PMF $(j,k)$ is indicated by $r^{(j)}_{k,n}\in\{0,1\}$, where $r^{(j)}_{k,n}=0$ means absence and $r^{(j)}_{k,n}=1$ means presence. We denote its position by $\boldsymbol{x}^{(j)}_{k,n}$ and define the augmented state $\boldsymbol{y}^{(j)}_{k,n}\triangleq\big[\boldsymbol{x}^{(j)\mathrm{T}}_{k,n},\,r^{(j)}_{k,n}\big]^{\mathrm{T}}$. We stack per-BS states as $\boldsymbol{y}^{(j)}_n\triangleq\big[\boldsymbol{y}^{(j)\mathrm{T}}_{1,n}\cdots\boldsymbol{y}^{(j)\mathrm{T}}_{K^{(j)}_{n,\mathrm{bi}},n}\big]^{\mathrm{T}}$ and all-BS states as $\boldsymbol{y}_n \triangleq \big[ \boldsymbol{y}^{(1)\mathrm{T}}_n \cdots \boldsymbol{y}^{(J)\mathrm{T}}_n \big]^{\mathrm{T}}$. Similarly, monostatic PMFs are indexed by $k\in {\mathcal{K}}^{(j)}_{n,\mathrm{mo}}\triangleq\{1,\ldots,K^{(j)}_{n,\mathrm{mo}}\}$, with augmented state $\tilde{\boldsymbol{y}}^{(j)}_{k,n}\triangleq\big[\tilde{\boldsymbol{x}}^{(j)\mathrm{T}}_{k,n},\,\tilde{r}^{(j)}_{k,n}\big]^{\mathrm{T}}$ (where the tilde denotes monostatic quantities), and stacked as $\tilde{\boldsymbol{y}}^{(j)}_n$ and $\tilde{\boldsymbol{y}}_n$ analogously.

At any time $n$, each PMF is either a legacy PMF (established in the past) or a new PMF (hypothesized at time $n$). 
%\footnote{Here, ``new PMF'' denotes a \emph{hypothesized} potential feature introduced for hypothesis generation and data association, \emph{not} an actually detected new target. Each PMF carries an existence indicator $r\in\{0,1\}$; a new PMF is regarded as \emph{confirmed newly detected} only when $r=1$.} 
For BS $j$ on the bistatic link, legacy and new PMFs are denoted by underline and overline respectively:
$\underline{\boldsymbol{y}}^{(j)}_{k,n}\triangleq
\big[\underline{\boldsymbol{x}}^{(j)\!\mathrm{T}}_{k,n},\,\underline{r}^{(j)}_{k,n}\big]^{\!\mathrm{T}}$ for $k\in\underline{\mathcal{K}}^{(j)}_{n,\mathrm{bi}}\triangleq\{1,\ldots,\underline{K}^{(j)}_{n,\mathrm{bi}}\}$ and
$\overline{\boldsymbol{y}}^{(j)}_{m,n}\triangleq
\big[\overline{\boldsymbol{x}}^{(j)\!\mathrm{T}}_{m,n},\,\overline{r}^{(j)}_{m,n}\big]^{\!\mathrm{T}}$ for $m\in\overline{\mathcal{K}}^{(j)}_{n,\mathrm{bi}}\triangleq\{1,\ldots,\overline{K}^{(j)}_{n,\mathrm{bi}}\}$.
Let $\mathcal{M}^{(j)}_{n,\mathrm{bi}}$ be the measurement index set with cardinality $M^{(j)}_{n,\mathrm{bi}}\triangleq\big|\mathcal{M}^{(j)}_{n,\mathrm{bi}}\big|$. \textcolor{black}{Under the single-bounce facade-reflection assumption, each bistatic measurement can be associated with at most one new PMF. Therefore, we set
$\overline{\mathcal{K}}^{(j)}_{n,\mathrm{bi}} \triangleq \mathcal{M}^{(j)}_{n,\mathrm{bi}}$
and thus
$\overline{K}^{(j)}_{n,\mathrm{bi}} = M^{(j)}_{n,\mathrm{bi}}$.}
The stacked vectors are $\underline{\boldsymbol{y}}^{(j)}_n \triangleq 
\big[\underline{\boldsymbol{y}}^{(j)}_{1,n},\ldots,\underline{\boldsymbol{y}}^{(j)}_{\underline{K}^{(j)}_{n,\mathrm{bi}},\,n}\big]$ and
$\overline{\boldsymbol{y}}^{(j)}_n \triangleq 
\big[\overline{\boldsymbol{y}}^{(j)}_{1,n},\ldots,\overline{\boldsymbol{y}}^{(j)}_{\overline{K}^{(j)}_{n,\mathrm{bi}},\,n}\big]$.
The monostatic case follows analogously: legacy and new states are
$\tilde{\underline{\boldsymbol{y}}}^{(j)}_{k,n}$ and
$\overline{\tilde{\boldsymbol{y}}}^{(j)}_{m,n}$ with measurement count $M^{(j)}_{n,\mathrm{mo}}\triangleq\big|\mathcal{M}^{(j)}_{n,\mathrm{mo}}\big|$ satisfying $\overline{K}^{(j)}_{n,\mathrm{mo}}=M^{(j)}_{n,\mathrm{mo}}$, and stacked as $\tilde{\underline{\boldsymbol{y}}}^{(j)}_{n}$ and $\overline{\tilde{\boldsymbol{y}}}^{(j)}_{n}$.

\subsection{State Transition Model}
This subsection specifies how the legacy-PMF sets at time $n$ are formed under
Schemes~I--II in Sec.~IV, and gives the corresponding state-transition kernels.

Scheme~I selects one link as \emph{dominant}.
The dominant link carries its entire PMFs set from time $n{-}1$ as legacy features
and augments it with the current measurement set, while the non-dominant link
enters only through its measurement likelihood.
For BS~$j$ with the bistatic link being dominant, the legacy set at time $n$ is
$
\underline{\mathcal{K}}^{(j)}_{n,\mathrm{bi}}=\mathcal{K}^{(j)}_{n-1,\mathrm{bi}},
\,
\mathcal{K}^{(j)}_{n,\mathrm{bi}}
=\underline{\mathcal{K}}^{(j)}_{n,\mathrm{bi}}\cup\mathcal{M}^{(j)}_{n,\mathrm{bi}},
$
and hence $\underline{K}^{(j)}_{n,\mathrm{bi}}=K^{(j)}_{n-1,\mathrm{bi}}$ and
$K^{(j)}_{n,\mathrm{bi}}=\underline{K}^{(j)}_{n,\mathrm{bi}}+M^{(j)}_{n,\mathrm{bi}}$.
The monostatic-dominant case follows by replacing the subscript $\mathrm{bi}$ with $\mathrm{mo}$.
The legacy PMF states evolve independently over time as
\begin{equation}f\left(\underline{\boldsymbol{y}}_n \middle| \boldsymbol{y}_{n-1} \right)=\prod_{j=1}^{J}\prod_{k=1}^{\underline{K}_{n,\mathrm{bi}}^{(j)}}f\left(\underline{\boldsymbol{y}}_{k,n}^{(j)}|\boldsymbol{y}_{k,n-1}^{(j)}\right), \label{eq:state1}\end{equation}
where $f\left(\underline{\boldsymbol{y}}_{k,n}^{(j)} \middle| \boldsymbol{y}_{k,n-1}^{(j)}\right) = f\left({\underline{\boldsymbol{x}}}_{k,n}^{(j)},\underline{r}_{k,n}^{(j)} \middle| \boldsymbol{x}_{k,n-1}^{(j)},r_{k,n-1}^{(j)}\right)$. 
If PMF $(j,k)$ exists at time $n{-}1$ (i.e., $r_{k,n-1}^{(j)}=1$), it either dies ($\underline{r}_{k,n}^{(j)}=0$) or survives ($\underline{r}_{k,n}^{(j)}=1$) at time $n$ with survival probability $P_{\mathrm{s}}$. If it survives, its position evolves according to $f\left(\underline{\boldsymbol{x}}_{k,n}^{(j)} \middle| \boldsymbol{x}_{k,n-1}^{(j)}\right)$. If $r_{k,n-1}^{(j)}=0$, the PMF cannot become a legacy feature at time $n$ (i.e., $\underline{r}_{k,n}^{(j)}=1$ has zero probability). The complete state transition is given by
\begin{equation}
\begin{aligned}
f&\left(\underline{\boldsymbol{x}}_{k,n}^{(j)},\underline{r}_{k,n}^{(j)} \middle| \boldsymbol{x}_{k,n-1}^{(j)}, r_{k,n-1}^{(j)}\right) \\
&= 
\begin{cases}
(1-P_{\mathrm{s}})f_{\mathrm{D}}\left(\underline{\boldsymbol{x}}_{k,n}^{(j)}\right), & r_{k,n-1}^{(j)}=1,\;\underline{r}_{k,n}^{(j)}=0, \\[3pt]
P_{\mathrm{s}}f\left(\underline{\boldsymbol{x}}_{k,n}^{(j)}|\boldsymbol{x}_{k,n-1}^{(j)}\right), & r_{k,n-1}^{(j)}=1,\;\underline{r}_{k,n}^{(j)}=1, \\[6pt]
f_{\mathrm{D}}\left(\underline{\boldsymbol{x}}_{k,n}^{(j)}\right), & r_{k,n-1}^{(j)}=0,\;\underline{r}_{k,n}^{(j)}=0, \\[3pt]
0, & r_{k,n-1}^{(j)}=0,\;\underline{r}_{k,n}^{(j)}=1,
\end{cases}
\label{eq:state_transition_scheme1}
\end{aligned}
\end{equation}
where $f_{\mathrm{D}}(\cdot)$ is an arbitrary dummy PDF.

Scheme~II processes the two links at each epoch $n$ in a fixed order.
The first link takes the other link's PMF set at $n{-}1$ as legacy, updates it using its
current measurements, and the second link then treats the updated set as its legacy and
augments it with its own measurements.
For the ordering $\mathrm{bi}\!\rightarrow\!\mathrm{mo}$, the set updates are $\underline{\mathcal{K}}^{(j)}_{n,\mathrm{bi}}=\mathcal{K}^{(j)}_{n-1,\mathrm{mo}},\,
\mathcal{K}^{(j)}_{n,\mathrm{bi}}
=\underline{\mathcal{K}}^{(j)}_{n,\mathrm{bi}}\cup\mathcal{M}^{(j)}_{n,\mathrm{bi}},\underline{K}^{(j)}_{n,\mathrm{bi}}=K^{(j)}_{n-1,\mathrm{mo}},\,
K^{(j)}_{n,\mathrm{bi}}
=\underline{K}^{(j)}_{n,\mathrm{bi}}+M^{(j)}_{n,\mathrm{bi}},$ followed by $\underline{\mathcal{K}}^{(j)}_{n,\mathrm{mo}}=\mathcal{K}^{(j)}_{n,\mathrm{bi}},\,
\mathcal{K}^{(j)}_{n,\mathrm{mo}}
=\underline{\mathcal{K}}^{(j)}_{n,\mathrm{mo}}\cup{\mathcal{M}}^{(j)}_{n,\mathrm{mo}},
\underline{K}^{(j)}_{n,\mathrm{mo}}=K^{(j)}_{n,\mathrm{bi}},\,
K^{(j)}_{n,\mathrm{mo}}
=\underline{K}^{(j)}_{n,\mathrm{mo}}+{M}^{(j)}_{n,\mathrm{mo}}.$ The reverse ordering ($\mathrm{mo}{\rightarrow}\mathrm{bi}$) is obtained by swapping link labels.
\begin{comment}
\begin{equation}\label{eq:seq-bi}
\begin{split}
\underline{\mathcal{K}}^{(j)}_{n,\mathrm{bi}}&=\mathcal{K}^{(j)}_{n-1,\mathrm{mo}},\quad
\mathcal{K}^{(j)}_{n,\mathrm{bi}}
=\underline{\mathcal{K}}^{(j)}_{n,\mathrm{bi}}\cup\mathcal{M}^{(j)}_{n,\mathrm{bi}},\\
\underline{K}^{(j)}_{n,\mathrm{bi}}&=K^{(j)}_{n-1,\mathrm{mo}},\quad
K^{(j)}_{n,\mathrm{bi}}
=\underline{K}^{(j)}_{n,\mathrm{bi}}+M^{(j)}_{n,\mathrm{bi}},
\end{split}
\end{equation}
followed by
\begin{equation}\label{eq:seq-mo}
\begin{split}
\underline{\mathcal{K}}^{(j)}_{n,\mathrm{mo}}&=\mathcal{K}^{(j)}_{n,\mathrm{bi}},\quad
\mathcal{K}^{(j)}_{n,\mathrm{mo}}
=\underline{\mathcal{K}}^{(j)}_{n,\mathrm{mo}}\cup\widetilde{\mathcal{M}}^{(j)}_{n,\mathrm{mo}},\\
\underline{K}^{(j)}_{n,\mathrm{mo}}&=K^{(j)}_{n,\mathrm{bi}},\quad
K^{(j)}_{n,\mathrm{mo}}
=\underline{K}^{(j)}_{n,\mathrm{mo}}+\widetilde{M}^{(j)}_{n,\mathrm{mo}}.
\end{split}
\end{equation}
\end{comment}

To model view-dependent cross-link visibility, Scheme~II introduces cross-link transitions
in addition to temporal survival, using the cross-link persistence probability $P_{\mathrm{c}}\in[0,1]$
and cross-link birth probability $P_{\mathrm{b}}\in[0,1]$.

\emph{Stage~1 ($\mathrm{mo}$ at $n{-}1$ to $\mathrm{bi}$ at $n$):}
Legacy states on the bistatic link evolve as
\begin{equation}
f\!\left(\underline{\boldsymbol{y}}_{n}\mid \tilde{\boldsymbol{y}}_{n-1}\right)
=\prod_{j=1}^{J}\prod_{k=1}^{\underline{K}^{(j)}_{n,\mathrm{bi}}}
f\!\left(\underline{\boldsymbol{y}}^{(j)}_{k,n}\mid \tilde{\boldsymbol{y}}^{(j)}_{k,n-1}\right),
\label{eq:scheme2_stage1_fact}
\end{equation}
where
\begin{equation}\label{eq:stage1-kernel}
\begin{split}
&f\!\left(\underline{\boldsymbol{x}}^{(j)}_{k,n},\underline{r}^{(j)}_{k,n}\mid
\tilde{\boldsymbol{x}}^{(j)}_{k,n-1},\tilde{r}^{(j)}_{k,n-1}\right)\\[2pt]
&\quad=
\begin{cases}
(1-P_{\mathrm{s}}P_{\mathrm{c}})\,f_{\mathrm{D}}(\underline{\boldsymbol{x}}^{(j)}_{k,n}),
& \tilde r^{(j)}_{k,n-1}=1,\;\underline r^{(j)}_{k,n}=0,\\[3pt]
P_{\mathrm{s}}P_{\mathrm{c}}\;f(\underline{\boldsymbol{x}}^{(j)}_{k,n}\mid \tilde{\boldsymbol{x}}^{(j)}_{k,n-1}),
& \tilde r^{(j)}_{k,n-1}=1,\;\underline r^{(j)}_{k,n}=1,\\[6pt]
(1-P_{\mathrm{b}})\,f_{\mathrm{D}}(\underline{\boldsymbol{x}}^{(j)}_{k,n}),
& \tilde r^{(j)}_{k,n-1}=0,\;\underline r^{(j)}_{k,n}=0,\\[3pt]
P_{\mathrm{b}}\;f_{\mathrm{n}}(\underline{\boldsymbol{x}}^{(j)}_{k,n}),
& \tilde r^{(j)}_{k,n-1}=0,\;\underline r^{(j)}_{k,n}=1.
\end{cases}
\end{split}
\end{equation}

\emph{Stage~2 ($\mathrm{bi}$ at $n$ to $\mathrm{mo}$ at $n$):}
The monostatic legacy inherits from the updated bistatic set as
\begin{equation}
f\!\left(\underline{\tilde{\boldsymbol{y}}}_{n}\mid \boldsymbol{y}_{n}\right)
=\prod_{j=1}^{J}\prod_{k=1}^{\underline{K}^{(j)}_{n,\mathrm{mo}}}
f\!\left(\underline{\tilde{\boldsymbol{y}}}^{(j)}_{k,n}\mid \boldsymbol{y}^{(j)}_{k,n}\right),
\label{eq:scheme2_stage2_fact}
\end{equation}
where
\begin{equation}\label{eq:stage2-kernel}
\begin{split}
&f\!\left(\tilde{\boldsymbol{x}}^{(j)}_{k,n},\tilde{r}^{(j)}_{k,n}\mid
\boldsymbol{x}^{(j)}_{k,n},r^{(j)}_{k,n}\right)\\[2pt]
&\quad=
\begin{cases}
(1-P_{\mathrm{c}})\,f_{\mathrm{D}}(\tilde{\boldsymbol{x}}^{(j)}_{k,n}),
& r^{(j)}_{k,n}=1,\;\tilde r^{(j)}_{k,n}=0,\\[3pt]
P_{\mathrm{c}}\;f(\tilde{\boldsymbol{x}}^{(j)}_{k,n}\mid \boldsymbol{x}^{(j)}_{k,n}),
& r^{(j)}_{k,n}=1,\;\tilde r^{(j)}_{k,n}=1,\\[6pt]
(1-P_{\mathrm{b}})\,f_{\mathrm{D}}(\tilde{\boldsymbol{x}}^{(j)}_{k,n}),
& r^{(j)}_{k,n}=0,\;\tilde r^{(j)}_{k,n}=0,\\[3pt]
P_{\mathrm{b}}\;f_{\mathrm{n}}(\tilde{\boldsymbol{x}}^{(j)}_{k,n}),
& r^{(j)}_{k,n}=0,\;\tilde r^{(j)}_{k,n}=1.
\end{cases}
\end{split}
\end{equation}

\noindent
Here, $f_{\mathrm{n}}(\cdot)$ is the birth prior. 
The deterministic setting is recovered with $P_{\mathrm{c}}{=}1$ and $P_{\mathrm{b}}{=}0$. 
The reverse ordering ($\mathrm{mo}{\rightarrow}\mathrm{bi}$) follows by interchanging link labels and corresponding state-transition PDFs.

\subsection{Likelihood Function Model}
\textcolor{black}{
This subsection specifies the likelihood functions for the bistatic and monostatic sensing views using the measurement and geometric relations in Sec.~II.}

\textcolor{black}{
For BS~$j$’s bistatic link, we use pseudo-range measurements. For a PMF state $\boldsymbol{x}^{(j)}_{k,n}$, define the predicted specular pseudo-range as
$
\mu^{(j)}_{k,n}\triangleq\|{\mathbf u}_n-\boldsymbol{x}^{(j)}_{k,n}\|.
$
As established in Sec.~II-B, the pseudo-range-domain estimator error is Gaussian with variance $\sigma_z^2=\mathrm c^2\sigma_\tau^2$.
For a specular component, the likelihood is
\begin{equation}
f_{\mathrm{spec}}\!\left(z^{(j)}_{m,n}\mid \boldsymbol{x}^{(j)}_{k,n}\right)
=\mathcal{N}\!\Big(z^{(j)}_{m,n};\,\mu^{(j)}_{k,n},\,\sigma_z^2\Big).
\end{equation}
For a diffuse subpath, the excess pseudo-range is modeled as
$V^{(j)}_{z,m,n}\sim\mathcal U[0,\psi_{z,n}]$, yielding
\begin{equation}
\begin{aligned}
f_{\mathrm{diff}}\!\big(z^{(j)}_{m,n}\mid \boldsymbol{x}^{(j)}_{k,n}\big)
= \int
  {\mathcal N}\!\big(z^{(j)}_{m,n};\,\mu^{(j)}_{k,n}+V^{(j)}_{z,m,n},\,\sigma_z^{2}\big)\,\\
  \times{\mathcal U}\!\big(V^{(j)}_{z,m,n};\,0,\,\psi_{z,n}\big)\,
  \mathrm{d}V^{(j)}_{z,m,n}.
\end{aligned}
\end{equation}
\noindent More generally, \(\mathcal  U(\cdot) \) can be replaced by a measurement-calibrated \( f_V(\cdot) \) for the excess pseudo-range \( V \), yielding the same marginal-likelihood form.}

\textcolor{black}{
For the monostatic link, the geometric consistency with the hypothesized reflector plane induced by $\boldsymbol{x}^{(j)}_{k,n}$ is quantified by the projection residual defined in Sec.~II-C. Let $\Pi^{(j)}_{k,n}$ denote the reflector plane induced by $\boldsymbol{x}^{(j)}_{k,n}$, and let the realized residual be $\boldsymbol e^{(j)}_{m,n,k}$ in \eqref{eq:mono_proj_residual}. Under the Gaussian-equivalent pseudo-position uncertainty model in Sec.~II-B and the residual propagation in Sec.~II-C, the residual covariance is $\boldsymbol{\Sigma}^{(j)}_{e,m,n,k}$ given in Sec.~II-C. Therefore, the monostatic likelihood is
\begin{equation}
f_{\mathrm{mo}}\!\left(\tilde{\boldsymbol z}^{(j)}_{m,n}\mid \tilde {\boldsymbol{x}}^{(j)}_{k,n}\right)
=
\mathcal{N}\!\left(
\boldsymbol e^{(j)}_{m,n,k};\,\mathbf 0,\,
\boldsymbol{\Sigma}^{(j)}_{e,m,n,k}
\right).
\label{eq:mono_like}
\end{equation}}

\subsection{Data Association Model}
The MPC measurements  $\boldsymbol{z}_{m,n}^{(j)}$ and $\tilde{\boldsymbol{z}}_{m,n}^{(j)}$, are subject to data association uncertainty. It is unknown which measurement corresponds to which PMF-$k$, and both false alarms and missed detections may occur.  For BS $j$ at time $n$, the bistatic-side and monostatic-side association vectors are defined respectively as
$\boldsymbol{b}_{n}^{(j)}=[{b}_{1,n}^{(j)},\ldots,
{b}_{M_{n,\mathrm{bi}}^{(j)},n}^{(j)}]$ and 
${c}_{n}^{(j)}=[{c}_{1,n}^{(j)},\ldots,
{c}_{{M}_{n,\mathrm{mo}}^{(j)},n}^{(j)}]$, and we collect them across all BSs as $\boldsymbol{b}_{n}=[\boldsymbol{b}_{n}^{(1)\mathrm{T}},\ldots,
\boldsymbol{b}_{n}^{(J)\mathrm{T}}]^{\mathrm{T}}$ and 
$\boldsymbol{c}_{n}=[\boldsymbol{c}_{n}^{(1)\mathrm{T}},\ldots,
\boldsymbol{c}_{n}^{(J)\mathrm{T}}]^{\mathrm{T}}$.
In Scheme~I, the association variable for the $l$th measurement on the bistatic-leading link adheres to the measurement-oriented model as \cite{Wielandner2023NonIdealSurfaces,Wielandner2024MIMO_NonIdealSurfaces,9585528}
\begin{equation}
\boldsymbol{b}_{l,n}^{(j)}\triangleq\begin{cases}k\in\{1,\ldots,\underline{K}_{n,\mathrm{bi}}^{(j)}+l\},&\text{if measurement \(l\) was}\\ &\text{generated by MF \(k\),}\\ 0,&\text{otherwise.}\end{cases}
\label{eq:DA_leading}
\end{equation}
The measurements from the auxiliary monostatic link can only originate from existing PMFs, and consequently, their association variables take values in the set $\{1, \ldots, K_{n,\mathrm{bi}}^{(j)}\}$ without generating new-feature hypotheses.  When the monostatic link acts as the leading mode, the roles of $\boldsymbol{b}_{n}^{(j)}$ and $\boldsymbol{c}_{n}^{(j)}$ are interchanged.
In Scheme~II, the bistatic and monostatic links play symmetric roles in an sequential update process, and all association variables follow the 
measurement-oriented model in~\eqref{eq:DA_leading} with the same interpretation of legacy and newly born PMFs. This representation of data association
makes it possible to develop scalable sum-product inference for environment mapping.

\section{Problem Formulation and Data Fusion}
This section formulates the environment mapping problem within a Bayesian detection and estimation framework. Two fusion schemes are introduced for combining monostatic and bistatic data. Based on the models in Sections II and III, we derive the joint posterior PDF and the corresponding factor graph. \textcolor{black}{For clarity, the overall inference and fusion pipeline, from pseudo-measurements to map outputs, is summarized in Workflow~\ref{alg:overall-flow}.}

\textcolor{black}{
\begin{algorithm}[t]
\caption{\textcolor{black}{Online inference and fusion pipeline (per BS).}}
\label{alg:overall-flow}
\textcolor{black}{%
\begin{algorithmic}[1]
\STATE Initialize the PMF set and prior beliefs at $n=0$.
\FOR{$n=1$ to $T$}
\STATE Acquire bistatic/monostatic MPC parameter estimates and their covariances, and convert them to pseudo-measurements $\boldsymbol z_n^{(j)}$ and $\tilde{\boldsymbol z}_n^{(j)}$ (Sec.~II).
\STATE Construct the factor-graph representation of the joint posterior over PMF (VA) states conditioned on the pseudo-measurements from both links (Secs.~III--IV).
\STATE Run SPA message passing on the factor graph: Scheme~I (unified graph, parallel update) or Scheme~II (sequential cross-link update) (Sec.~IV-D).
\STATE For each maintained PMF $(j,k)$, compute the marginal posterior belief and existence probability $\Pr(r^{(j)}_{k,n}=1|\cdot)$; prune PMFs with $\Pr(r^{(j)}_{k,n}=1|\cdot)<P_{\mathrm{prun}}$, and compute MMSE location estimates from the marginal posteriors.
\ENDFOR
\STATE Aggregate the per-BS confirmed PMFs/VAs (with $\Pr(r^{(j)}_{k,n}=1|\cdot)\ge P_{\mathrm{th}}$) and their MMSE location estimates $\hat{\boldsymbol{x}}^{(j)}_{k,n}$ across $j=1,\ldots,J$ to form the overall map.
\end{algorithmic}
}
\end{algorithm}
}

\subsection{Problem Formulation}
Given the measurements from both links, \( \boldsymbol{z}_{1:N} \triangleq [\boldsymbol{z}_1^{\mathrm{T}}, \dots, \boldsymbol{z}_N^{\mathrm{T}}]^{\mathrm{T}} \) and \( \tilde{\boldsymbol{z}}_{1:N} \), where each \( \boldsymbol{z}_n \) is the set of measurements from the \( J \) BSs at epoch \( n \), i.e., \( \boldsymbol{z}_n \triangleq [\boldsymbol{z}_n^{(1)\mathrm{T}}, \dots, \boldsymbol{z}_n^{(J)\mathrm{T}}]^{\mathrm{T}} \), our goal is to estimate the PMF states associated with the \( J \) BSs over the \( N \) epochs. \textcolor{black}{We perform inference independently for each BS (i.e., no inter-BS message passing), and the overall map is obtained by aggregating the per-BS estimates.}
A PMF $(j,k)$ is declared to exist at epoch \( n \) if its posterior existence probability exceeds a threshold:
$
\Pr(r_{k,n}^{(j)}=1 \mid \boldsymbol{z}_{1:n}, \tilde{\boldsymbol{z}}_{1:n}) > P_{\mathrm{th}},
$
where this probability is obtained by marginalizing the posterior of the augmented state
\( f(\boldsymbol{x}_{k,n}^{(j)}, r_{k,n}^{(j)} \mid \boldsymbol{z}_{1:n}, \tilde{\boldsymbol{z}}_{1:n}) \).
For detected PMFs, the location is estimated by the MMSE estimator~\cite{kay1993fundamentals}:
$
\hat{\boldsymbol{x}}_{k,n}^{(j)\mathrm{MMSE}} \triangleq \mathbb{E}\left[\boldsymbol{x}_{k,n}^{(j)} \mid r_{k,n}^{(j)}=1, \boldsymbol{z}_{1:n}, \tilde{\boldsymbol{z}}_{1:n}\right].
$
To prevent unbounded growth of the hypothesis set, we prune PMFs whose existence probability falls below \( P_{\mathrm{prun}} \). 
The same procedure applies to the monostatic-link PMFs (tilde quantities).

\subsection{Joint Posterior PDF for Scheme I}
We now derive the joint PDF function for the first proposed fusion mechanism, Scheme~I, which operates under a single-link-dominant paradigm.  Formally, we model the posterior distribution within a unified Bayesian framework \footnote{We use $f(\cdot\mid\cdot\,;\,\cdot)$ to separate random conditioning variables from known inputs; here $\mathbf u_{n'}$ is observed (non-random) and thus appears after the semicolon.} as  
\begin{equation}
 \begin{aligned}
f\big(&\boldsymbol{y}_{0:n}^{*}, \boldsymbol{b}_{1:n}, \boldsymbol{c}_{1:n},
  \boldsymbol{m}_{1:n}, \tilde{\boldsymbol{m}}_{1:n} \mid 
  \boldsymbol{z}_{1:n}, \tilde{\boldsymbol{z}}_{1:n}; \mathbf{u}_{0:n}\big)  \\
& \propto f\big(\boldsymbol{y}_{0:n}^{*}, \boldsymbol{b}_{1:n}, \boldsymbol{c}_{1:n},
  \boldsymbol{m}_{1:n}, \tilde{\boldsymbol{m}}_{1:n}; \mathbf{u}_{0:n}\big) \\
& \quad \times f\big(\boldsymbol{z}_{1:n}, \tilde{\boldsymbol{z}}_{1:n} \mid
  \boldsymbol{y}_{0:n}^{*}, \boldsymbol{b}_{1:n}, \boldsymbol{c}_{1:n},
  \boldsymbol{m}_{1:n}, \tilde{\boldsymbol{m}}_{1:n}; \mathbf{u}_{0:n} \big)  
  \label{eq:bys},
  \end{aligned}
\end{equation}
where $\boldsymbol{m}_{1:n}\triangleq[(\boldsymbol{m}_{1:n}^{(1)})^{\mathrm T},\ldots,(\boldsymbol{m}_{1:n}^{(J)})^{\mathrm T}]^{\mathrm T}$ and
$\tilde{\boldsymbol{m}}_{1:n}\triangleq[(\tilde{\boldsymbol{m}}_{1:n}^{(1)})^{\mathrm T},\ldots,(\tilde{\boldsymbol{m}}_{1:n}^{(J)})^{\mathrm T}]^{\mathrm T}$ stack the per-BS measurement-count sequences over time, with
$\boldsymbol{m}_{1:n}^{(j)}=[m_1^{(j)},\ldots,m_n^{(j)}]^{\mathsf T}$ and
$\tilde{\boldsymbol{m}}_{1:n}^{(j)}=[\tilde m_1^{(j)},\ldots,\tilde m_n^{(j)}]^{\mathsf T}$ denoting the bistatic and monostatic counts, respectively.
Before observation, $m_n^{(j)}$ and $\tilde m_n^{(j)}$ are cardinality random variables; after extracting the MPC sets at epoch $n$, their realizations are the observed counts
$M_{n,\mathrm{bi}}^{(j)}\triangleq|\mathcal{M}_{n,\mathrm{bi}}^{(j)}|$ and
$M_{n,\mathrm{mo}}^{(j)}\triangleq|\mathcal{M}_{n,\mathrm{mo}}^{(j)}|$.
Accordingly, throughout inference we condition on these observed counts (equivalently, on the measurement sets), and keep $m_n^{(j)}$ and $\tilde m_n^{(j)}$ explicit only to match the standard factorization. Here, $\boldsymbol{y}_n^{\ast}$ depends on the scheme: under Scheme~I,
$\boldsymbol{y}_n^{\ast}\triangleq \boldsymbol{y}_n^{(\mathrm{dom})}$, where
$\boldsymbol{y}_n^{(\mathrm{dom})}\in\{\boldsymbol{y}_n,\tilde{\boldsymbol{y}}_n\}$ denotes the dominant-link state;
under Scheme~II,
$\boldsymbol{y}_n^{\ast}\triangleq[\boldsymbol{y}_n^{\mathrm T},\tilde{\boldsymbol{y}}_n^{\mathrm T}]^{\mathrm T}$.

Based on (\ref{eq:bys}) and under common assumptions \cite{Leitinger2019BP_SLAM},\,\,\cite{bar-shalom2011tracking}, the joint PDF for Scheme~I can be expressed as (illustrated with the bistatic link as the dominant example)
\begin{equation}
\begin{aligned}
f\big(&\boldsymbol{y}_{0:n}, \boldsymbol{b}_{1:n}, \boldsymbol{c}_{1:n},
  \boldsymbol{m}_{1:n}, \tilde{\boldsymbol{m}}_{1:n} \mid 
  \boldsymbol{z}_{1:n}, \tilde{\boldsymbol{z}}_{1:n}; \mathbf{u}_{0:n}\big)  \\
=& \prod_{j'=1}^{J} \prod_{k'=1}^{K_{0,\mathrm{bi}}^{(j')}} 
    \underbrace{f\big(\boldsymbol{y}_{k',0}^{(j')};\mathbf{u}_0\big)}_{(a)} 
    \prod_{n^{''}=1}^{n} \prod_{k=1}^{K_{n^{''}-1,\mathrm{bi}}^{(j')}} 
    \underbrace{f\big(\boldsymbol{y}_{k,n^{''}}^{(j')} \mid \boldsymbol{y}_{k,n^{''}-1}^{(j')}\big)}_{(b)}  \\
& \times \prod_{j=1}^{J} \prod_{n'=1}^{n} 
    \underbrace{f\big(\boldsymbol{b}_{n'}^{(j)}, m_{n'}^{(j)}, \overline{\boldsymbol{y}}_{n'}^{(j)} 
    \mid \boldsymbol{y}_{n'}^{(j)}; \mathbf{u}_{n'}\big)}_{(c)} 
    \underbrace{f\big(\boldsymbol{c}_{n'}^{(j)}, \tilde{m}_{n'}^{(j)} \mid 
    \boldsymbol{y}_{n'}^{(j)}; \mathbf{u}_{n'}\big)}_{(d)}  \\
& \times 
    \underbrace{f\big(\boldsymbol{z}_{n'}^{(j)} \mid \boldsymbol{y}_{n'}^{(j)}, 
    \boldsymbol{b}_{n'}^{(j)}, m_{n'}^{(j)}; \mathbf{u}_{n'}\big)}_{(e)} 
    \underbrace{f\big(\tilde{\boldsymbol{z}}_{n'}^{(j)} \mid \boldsymbol{y}_{n'}^{(j)}, 
    \boldsymbol{c}_{n'}^{(j)}, \tilde{m}_{n'}^{(j)}; \mathbf{u}_{n'}\big)}_{(f)},
    \label{eq:jointpdf}
\end{aligned}
\end{equation}
where (a) denotes the prior for the initial PMFs (and can be omitted if unavailable), and (b) denotes the state-transition PDF in Sec.~III.
Terms (c) and (d) are the bistatic/monostatic data-association factors, respectively, and (e) and (f) are the corresponding measurement-likelihood factors.
Terms (c) and (d) can be instantiated following~\cite{9585528} (Supplementary Material, Sec.~1.1).

The bistatic likelihood factor (e) admits the expanded form in \eqref{eq:LH1}, where
$\mathcal{M}_{\mathbf{b}_{n'}^{(j)},k}$ and $\mathcal{M}_{\mathbf{b}_{n'}^{(j)},\,k+\underline{K}_{n',\mathrm{bi}}}$
denote the index sets of measurements associated with the $k$th legacy PMF and the $k$th new PMF, respectively.
\begin{equation}
\begin{aligned}
f\big(&\boldsymbol{z}_{n'}^{(j)} \mid \boldsymbol{y}_{n'}^{(j)}, \boldsymbol{b}_{n'}^{(j)}, m_{n'}^{(j)}; \mathbf{u}_{n'}\big) \\
=& \prod_{l=1}^{m_{n^{\prime}}^{(j)}}f_{\mathrm{fa}}(z_{l,n^{\prime}}^{(j)})
\prod_{k=1}^{m_{n^{\prime}}^{(j)}}\prod_{l\in\mathcal{M}_{\mathbf{b}_{n^{\prime}}^{(j)},k+\underline{K}_{{n',\mathrm{bi}}}}}\frac{f({z}_{l,n^{\prime}}^{(j)}|\overline{\boldsymbol{x}}_{k,n^{\prime}}^{(j)};\mathbf{u}_{n^{\prime}})}{f_{\mathrm{fa}}({z}_{l,n^{\prime}}^{(j)})}\\
&\times \prod_{k=1}^{\underline{K}_{{n',\mathrm{bi}}}}\prod_{l\in\mathcal{M}_{\mathbf{b}_{n^{\prime}}^{(j)},k}}\frac{f({z}_{l,n^{\prime}}^{(j)}|\underline{\boldsymbol{x}}_{k,n^{\prime}}^{(j)};\mathbf{u}_{n^{\prime}})}{f_{\mathrm{fa}}({z}_{l,n^{\prime}}^{(j)})},
\label{eq:LH1}
\end{aligned}
\end{equation}
\textcolor{black}{where $f_{\mathrm{fa}}(\cdot)$ denotes the i.i.d.\ clutter (false-alarm) measurement PDF (instantiated as uniform over the corresponding measurement domain in our simulations).} Factor (f) is defined analogously to (e) using the corresponding monostatic quantities.
\vspace{-0.9mm}
%% 方案1单一视角主导型的因子分解如下：
\begin{figure*}[!b]
\centering
\vspace{-4.3mm}
\noindent\hrulefill
\vspace{-3pt} % 减少两个公式之间的间距
\begin{equation}\begin{aligned}
f\big(&\boldsymbol{y}_{0:n},\boldsymbol{b}_{1:n},\boldsymbol{c}_{1:n},\boldsymbol{m}_{1:n},\tilde{\boldsymbol{m}}_{1:n}\mid\boldsymbol{z}_{1:n},\tilde{\boldsymbol{z}}_{1:n};\mathbf{u}_{0:n}\big) \\
 \propto& 
\underbrace{\prod_{j'=1}^{J}\prod_{{k''}=1}^{K_{0,\mathrm{bi}}^{(j')}} f\left(\boldsymbol{y}_{{k''},0}^{(j')};\mathbf{u}_0\right)}_{\text{prior PDFs }}
\prod_{j=1}^{J} \prod_{n'=1}^{n} 
%%虚警
\underbrace{\left(\prod_{l=1}^{m_{n}^{(j)}}f_{\mathrm{fa}}\left({z}_{l,n^{\prime}}^{(j)}\right)\right)\left(\prod_{l=1}^{\tilde{m}_{n^{\prime}}^{(j)}}f_{\mathrm{fa}}\left(\tilde{\boldsymbol{z}}_{l,n^{\prime}}^{(j)}\right)\right)}_{\text{prior related to false alarms}} \\
%% 继承特征状态转移
& \times\prod_{k=1}^{\underline{K}_{n^\prime,\mathrm{bi}}^{(j)}} \underbrace{\underline{q}\left(\underline{\boldsymbol{x}}_{k,n'}^{(j)},\underline{r}_{k,n'}^{(j)} \big| \boldsymbol{x}_{k,n'-1}^{(j)},{r}_{k,n'-1}^{(j)}\right)}_{\text{legacy PMFs state transition}}
%% h_k
\prod_{l=1}^{m_{n'}^{(j)}}
\underbrace{
h_{k}\left(\underline{\boldsymbol{x}}_{k,n^{\prime}}^{(j)},\underline{r}_{k,n^{\prime}}^{(j)},b_{l,n^{\prime}}^{(j)};{z}_{l,n^{\prime}}^{(j)},\mathbf{u}_{n^{\prime}}\right)}_{\text{likelihood factor for leagacy PMFs (bistatic link) }}
%%\tilde{h}_k
\prod_{l=1}^{\tilde{m}_{n'}^{(j)}}
\underbrace{ 
\tilde{h}_{k}\left(\underline{\boldsymbol{x}}_{k,n^{\prime}}^{(j)},\underline{r}_{k,n^{\prime}}^{(j)},c_{l,n^{\prime}}^{(j)};\tilde{\boldsymbol{z}}_{l,n^{\prime}}^{(j)},\mathbf{u}_{n^{\prime}}\right)}_{\text{likelihood factor for leagacy PMFs (monostatic link) }}
\\
%% 新特征概率分布
&\times\prod_{k'=1}^{m_{n'}^{(j)}}
\underbrace{\overline{q}\left(\overline{\boldsymbol{x}}_{k',n'}^{(j)},\overline{r}_{k',n'}^{(j)};\mathbf{u}_{n'}\right)}_{\text{prior factor for new features}}
%% g_{1,K+k}
\underbrace{
{g}_{\underline{K}_{n',\mathrm{bi}}^{(j)}+k'}\left(\overline{\boldsymbol{x}}_{k',n^{\prime}}^{(j)},\overline{r}_{k',n^{\prime}}^{(j)},b_{k',n^{\prime}}^{(j)};{z}_{k',n^{\prime}}^{(j)},\mathbf{u}_{n^{\prime}}\right)}_{\text{factor node for new feature labeling}}
\\
&\times \prod_{l=k'+1}^{m_{n'}^{(j)}}  
\underbrace{
{h}_{\underline{K}_{n',\mathrm{bi}}^{(j)}+k'}\left(\overline{\boldsymbol{x}}_{k',n^{\prime}}^{(j)},\overline{r}_{k',n^{\prime}}^{(j)},b_{l,n^{\prime}}^{(j)};{z}_{l,n^{\prime}}^{(j)},\mathbf{u}_{n'}\right)}_{\text{likelihood factor for new PMFs (bistatic link) }}
\prod_{l=1}^{\tilde{m}_{n'}^{(j)}}
\underbrace{\tilde{h}_{\underline{K}_{n',\mathrm{bi}}^{(j)}+k'}\left(\overline{\boldsymbol{x}}_{k',n^{\prime}}^{(j)},\overline{r}_{k',n^{\prime}}^{(j)},c_{l,n^{\prime}}^{(j)};\tilde{\boldsymbol{z}}_{l,n^{\prime}}^{(j)},\mathbf{u}_{n'}\right)}_{\text{likelihood factor for new PMFs (monostatic link) }}\\
\label{eq:F1}
\end{aligned}\end{equation}
\noindent\hrulefill
\end{figure*}

Under common conditional-independence assumptions~\cite{Leitinger2019BP_SLAM,bar-shalom2011tracking}, the joint posterior in \eqref{eq:jointpdf} admits the factorization in \eqref{eq:F1}, and the corresponding factor-node definitions are given next.

%%下面的代码报错了
The distribution $\underline{q}\!\big(\underline{\boldsymbol{x}}_{k,n}^{(j)}, 
\underline{r}_{k,n}^{(j)} \mid \boldsymbol{x}_{k,n-1}^{(j)}, r_{k,n-1}^{(j)}\big)$ 
denotes the pseudo-legacy PMF state transition and is defined as follows
\begin{equation}
\begin{aligned}
\underline{q}\Big(&\underline{\boldsymbol{x}}_{k,n}^{(j)},\underline{r}_{k,n}^{(j)} \big| \boldsymbol{x}_{k,n-1}^{(j)},{r}_{k,n-1}^{(j)}\Big)
\\=
&\begin{cases}
e^{-\mu_m(\underline{\boldsymbol{x}}^{(j)}_{k,n};\mathbf{u}_n)}\,
f\!\left(\underline{\boldsymbol{x}}^{(j)}_{k,n},1\,\middle|\,\boldsymbol{x}^{(j)}_{k,n-1},r^{(j)}_{k,n-1}\right),
& \underline{r}^{(j)}_{k,n}=1,\\
f\!\left(\underline{\boldsymbol{x}}^{(j)}_{k,n},0\,\middle|\,\boldsymbol{x}^{(j)}_{k,n-1},r^{(j)}_{k,n-1}\right),
& \underline{r}^{(j)}_{k,n}=0,
\end{cases}
\label{eq:qdef}
\end{aligned}
\end{equation}
where $\mu_{\mathrm m}(\cdot)$ denote the Poisson mean numbers of MPCs generated by a facade on the bistatic link.

The distribution $\overline{q}\!\left(\overline{\boldsymbol{x}}_{k,n}^{(j)},\, 
\overline{r}_{k,n}^{(j)};\, \mathbf{u}_{n}\right)$ denotes the pseudo-probability 
distribution of the new PMFs as fowllows
\begingroup
\setlength\arraycolsep{3pt}
\begin{equation}
\begin{aligned}
\overline{q}\!\big(&\overline{\boldsymbol{x}}_{k,n}^{(j)},\,\overline{r}_{k,n}^{(j)};\,\mathbf{u}_{n}\big)\\
&\triangleq
\begin{cases}
\mu_{\mathrm{n}}\,
f_{\mathrm{n}}\!\big(\overline{\boldsymbol{x}}_{k,n}^{(j)};\,\mathbf{u}_{n}\big)\,
\displaystyle\frac{e^{-\mu_{\mathrm{m}}(\overline{\boldsymbol{x}}_{k,n}^{(j)};\,\mathbf{u}_{n})}}
{1-e^{-\mu_{\mathrm{m}}(\overline{\boldsymbol{x}}_{k,n}^{(j)};\,\mathbf{u}_{n})}}, 
& \overline{r}_{k,n}^{(j)}=1,\\[6pt]
f_{\mathrm{D}}\!\big(\overline{\boldsymbol{x}}_{k,n}^{(j)}\big), 
& \overline{r}_{k,n}^{(j)}=0.
\end{cases}
\end{aligned}
\end{equation}
\textcolor{black}{At each epoch, the number of newly born PMFs is Poisson with mean $\mu_{\mathrm n}$. Here, $f_{\mathrm n}(\cdot)$ is the birth PDF for initializing $\overline{\boldsymbol{x}}_{k,n}^{(j)}$.}

The function ${g}_{\underline{K}_{n,\mathrm{bi}}^{(j)}+k}\!\big(\cdot\big)$
denotes the pseudo-likelihood of the {first measurement} associated with a
{new PMF}. If the new PMF
exists, it  is as follows
% --- tighter cases and short lines for two-column layout ---
\begingroup
\setlength\arraycolsep{2pt}
\begin{align}
&g_{\underline{K}_{n,\mathrm{bi}}^{(j)}+k}\!\big(\overline{\boldsymbol{x}}_{k,n}^{(j)},
\overline{r}_{k,n}^{(j)}=1, b_{k,n}^{(j)};
{z}_{l,n}^{(j)}, \mathbf{u}_{n}\big)\notag\\[-2pt]
&\quad\triangleq
\begin{cases}
\dfrac{\mu_{\mathrm{m}}\!\big(\overline{\boldsymbol{x}}_{k,n}^{(j)};\mathbf{u}_{n}\big)
      f\!\big(\boldsymbol{z}_{l,n}^{(j)} \mid \overline{\boldsymbol{x}}_{k,n}^{(j)}; \mathbf{u}_{n}\big)}
     {\mu_{\mathrm{fa}}\, f_{\mathrm{fa}}\!\big(\boldsymbol{z}_{l,n}^{(j)}\big)},
& b_{k,n}^{(j)}=\underline{K}_{n,\mathrm{bi}}^{(j)}+k,\\[6pt]
0, & b_{k,n}^{(j)}\neq \underline{K}_{n,\mathrm{bi}}^{(j)}+k.
\end{cases}
\label{eq:g-new-exists}
\end{align}
\noindent \textcolor{black}{Here, $\mu_{\mathrm{fa}}$ denotes the mean number of false alarms per sensing epoch under the assumed Poisson clutter-cardinality model.}
\endgroup

If the new PMF does not exist ($\overline{r}^{(j)}_{k,n}=0$), we set
$g_{\underline{K}_{n}^{(j)}+k}(\cdot)=0$ when $b^{(j)}_{k,n'}=\underline{K}^{(j)}_{n,\mathrm{bi}}+k$ (i.e., the $k$th new-PMF label is selected), and $g_{\underline{K}_{n}^{(j)}+k}(\cdot)=1$ otherwise.

% 版本 4（强调第二个自变量取值 r_{k,n}^{(j)}=1）
When $r_{k,n}^{(j)}=0$, we define
$h_k(\boldsymbol{x}_{k,n}^{(j)},0,b_{l,n}^{(j)};z_{l,n}^{(j)},\mathbf u_n)=0$
if $b_{l,n}^{(j)}=k$, and
$h_k(\cdot)=1$ otherwise.
For $r_{k,n}^{(j)}=1$, the term
$h_k\!\big(\boldsymbol{x}_{k,n}^{(j)},1,b_{l,n}^{(j)};z_{l,n}^{(j)},\mathbf u_n\big)$
is defined as the general pseudo-likelihood
\begin{equation}
\begin{aligned}
h_k\!\Big(\boldsymbol{x}_{k,n}^{(j)},&,1,b_{l,n}^{(j)};z_{l,n}^{(j)},\mathbf u_n\Big)\\
&\triangleq
\begin{cases}
\dfrac{\mu_\mathrm{m}\!\left(\boldsymbol{x}_{k,n}^{(j)};\mathbf u_n\right)
\,f\!\left(z_{l,n}^{(j)} \mid \boldsymbol{x}_{k,n}^{(j)},\mathbf u_n\right)}
{\mu_\mathrm{fa}\,f_\mathrm{fa}\!\left(z_{l,n}^{(j)}\right)},
& b_{l,n}^{(j)}=k,\\[6pt]
1, & b_{l,n}^{(j)}\neq k.
\end{cases}
\end{aligned}
\label{genh}
\end{equation}
The terms
$\tilde{h}_{\underline{K}_{n}^{(j)}+k}\!\left(\overline{\boldsymbol{x}}_{k,n}^{(j)}\right)$,
$h_{\underline{K}_{n}^{(j)}+k}\!\left(\overline{\boldsymbol{x}}_{k,n}^{(j)}\right)$,
$h_k\!\left(\underline{\boldsymbol{x}}_{k,n}^{(j)}\right)$, and
$\tilde{h}_k\!\left(\underline{\boldsymbol{x}}_{k,n}^{(j)}\right)$ in (\ref{eq:F1})
all follow the same functional form as the general pseudo-likelihood in~\eqref{genh}.

\begin{figure*}[t]
    \centering
    \includegraphics[width=1\linewidth]{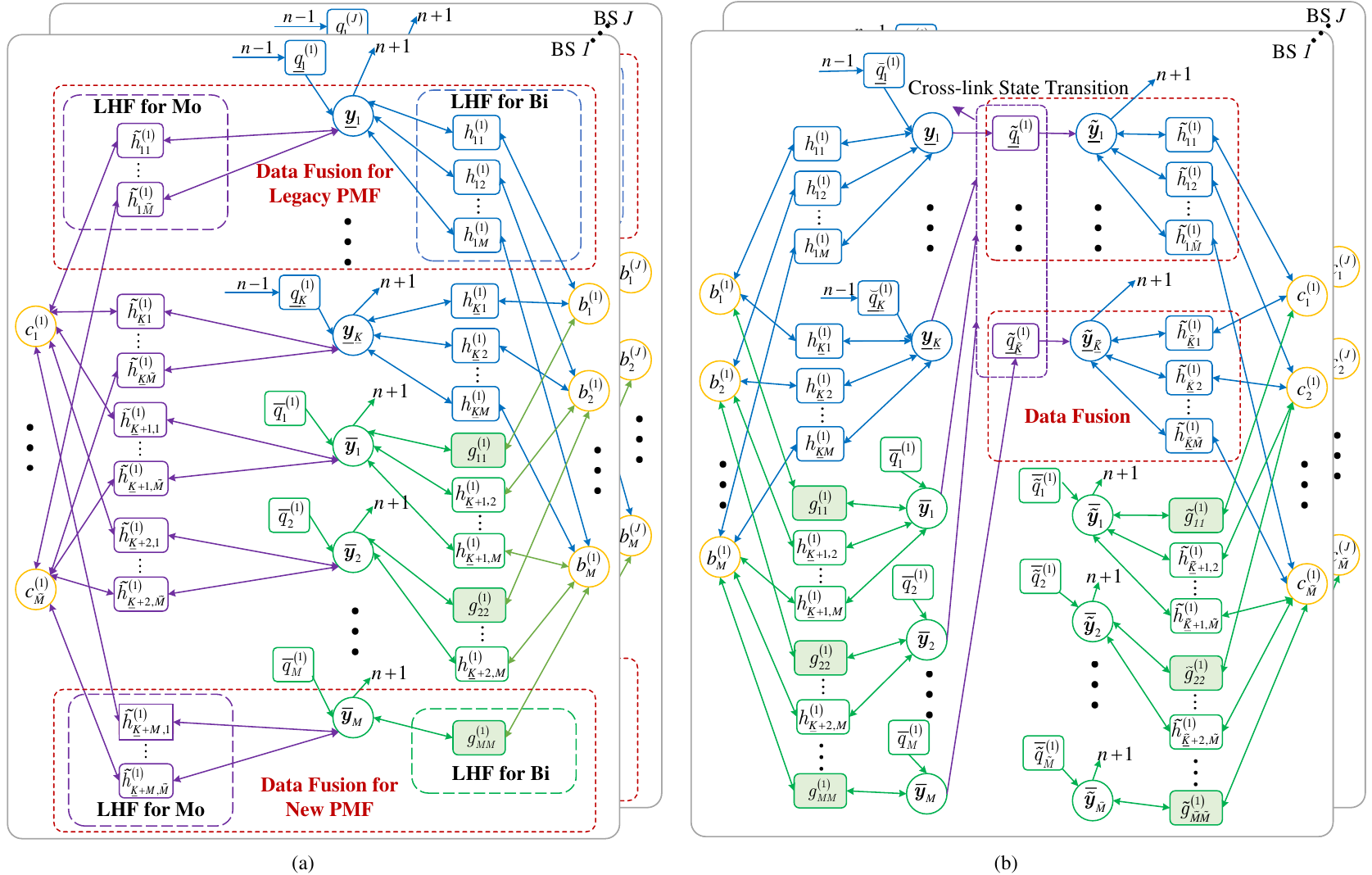}
    \caption{Factor graph representing the factorization of the joint posterior PDF of the proposed schemes at time instant $n$. For readability, the time index $n$ is omitted from all variables in the graphs. (a) Factor graph of the joint posterior PDF for Scheme I in (\ref{eq:F1}). (b) Factor graph of the joint posterior PDF for Scheme II in (\ref{eq:F2}). The sub-graphs corresponding to individual BSs are indicated by white rounded rectangles. Factor nodes are depicted as rectangular boxes, while variable nodes are represented as circles. For the factor graph corresponding to BS~$j$, the following shorthand notations are used:
    $\underline{K}\triangleq \underline{K}^{(j)}_{n,\mathrm{bi}}$,
    $\tilde{\underline{K}}\triangleq \underline{K}^{(j)}_{n,\mathrm{mo}}$, ${M}\triangleq {M}^{(j)}_{n,\mathrm{bi}}$,
    $\tilde{{M}}\triangleq {M}^{(j)}_{n,\mathrm{mo}}$, 
    $b_m^{(j)}\triangleq b_{m,n}^{(j)}$, 
    $c_m^{(j)}\triangleq c_{m,n}^{(j)}$, $\underline{q}_k^{(j)} \triangleq\underline{q}\!\big(\underline{\boldsymbol{x}}_{k,n}^{(j)}, \underline{r}_{k,n}^{(j)} \mid \boldsymbol{x}_{k,n-1}^{(j)}, r_{k,n-1}^{(j)}\big)$, 
    $\tilde{\underline{q}}_k^{(j)} \triangleq \tilde{\underline{q}}\!\big(\tilde{\underline{\boldsymbol{x}}}_{k,n}^{(j)}, 
    \tilde{\underline{r}}_{k,n}^{(j)} \mid {\boldsymbol{x}}_{k,n}^{(j)}, \tilde{r}_{k,n}^{(j)}\big)$, $\overset\smile{\underline{q}}_k^{(j)} \triangleq {\underline{q}}\!\big({\underline{\boldsymbol{x}}}_{k,n}^{(j)}, 
    {\underline{r}}_{k,n}^{(j)} \mid \tilde{\boldsymbol{x}}_{k,n-1}^{(j)}, \tilde{r}_{k,n-1}^{(j)}\big)$, 
    $\overline{q}_{m}^{(j)}\triangleq \overline{q}\!\big(\overline{\boldsymbol{x}}_{k,n}^{(j)},\,\overline{r}_{k,n}^{(j)};\,\mathbf{u}_{n}\big)$, 
    $\overline{\tilde{q}}_{m}^{(j)}\triangleq \overline{\tilde{q}}\!\big(\overline{\boldsymbol{\tilde{x}}}_{k,n}^{(j)},\,\overline{\tilde{r}}_{k,n}^{(j)};\,\mathbf{u}_{n}\big)$,
    $g_{kk}^{(j)} \triangleq g_{\underline{K}_{n}^{(j)}+k}\!\big(
    \overline{\boldsymbol{x}}_{k,n}^{(j)},
    \overline{r}_{k,n}^{(j)},\,
    b_{k,n}^{(j)};\,
   \mathrm{z}_{k,n}^{(j)}, \mathbf{u}_{n}
   \big)$ where $\underline{K}_{n}^{(j)} \triangleq \underline{K}^{(j)}_{n,\mathrm{bi}}$ in Fig.2 (a) and Fig.2 (b), 
   $\tilde{g}_{kk}^{(j)} \triangleq \tilde{g}_{\underline{K}_{n}^{(j)}+k}\!\big(
   \overline{\boldsymbol{\tilde{x}}}_{k,n}^{(j)},
   \overline{\tilde{r}}_{k,n}^{(j)},\,
   c_{k,n}^{(j)};\,
   \tilde{\boldsymbol{z}}_{k,n}^{(j)}, \mathbf{u}_{n}
   \big)$, 
   $h_{kl}^{(j)} \triangleq h_{\underline{K}+k,l}^{(j)}\triangleq h_k\!\Big(\boldsymbol{x}_{k,n}^{(j)}, r_{k,n}^{(j)}, b_{l,n}^{(j)}; {z}_{l,n}^{(j)}, \mathbf{u}_n\Big)$,
   $\tilde{h}_{kl}^{(j)} \triangleq \tilde{h}_{\underline{K}+k,l}^{(j)}\triangleq \tilde{h}_k\!\Big(\boldsymbol{x}_{k,n}^{(j)}, r_{k,n}^{(j)}, b_{l,n}^{(j)}; \tilde{z}_{l,n}^{(j)}, \mathbf{u}_n\Big)$.
   }
    \label{fig:factorPic}
    %\vspace{-8pt}
\end{figure*}
%\vspace{-18pt}
\subsection{Joint Posterior PDF for Scheme II}

Building on the single-link-dominant strategy in Scheme~I, we next introduce a cross-link sequential fusion scheme (Scheme~II) to better exploit the complementarity of the bistatic and monostatic links. Unlike Scheme~I, Scheme~II performs inference sequentially: one link first updates the posterior using its measurements, and the resulting posterior is then used as the prior for the other link. This stepwise procedure enables effective fusion of both sensing views without fully joint processing. The resulting joint posterior is given by
\begin{equation}
\begin{aligned}
 & f\!\left(\boldsymbol{y}_{0:n},\tilde{\boldsymbol{y}}_{0:n},\boldsymbol{b}_{1:n},\boldsymbol{c}_{1:n},\boldsymbol{m}_{1:n},\tilde{\boldsymbol{m}}_{1:n}\mid \boldsymbol{z}_{1:n},\tilde{\boldsymbol{z}}_{1:n}\,;\,\mathbf{u}_{0:n}\right) \\
 & \propto \prod_{j^{\prime}=1}^{J}\!\left(\prod_{k=1}^{K_{0,\mathrm{bi}}^{(j^{\prime})}}\underbrace{f\!\left(\boldsymbol{y}_{k,0}^{(j^{\prime})}\,;\,\mathbf{u}_{0}\right)}_{(a)}\right)\!
   \left(\prod_{k=1}^{{K}_{0,\mathrm{mo}}^{(j^{\prime})}}\underbrace{f\!\left(\tilde{\boldsymbol{y}}_{k^{\prime},0}^{(j^{\prime})}\,;\,\mathbf{u}_{0}\right)}_{(b)}\right) \\
 & \times \prod_{j=1}^{J}\prod_{n^{\prime}=1}^{n}
   \underbrace{f\!\left(\underline{\boldsymbol{y}}_{n^{\prime}}^{(j)} \,\big|\, \tilde{\boldsymbol{y}}_{n^{\prime}-1}^{(j)}\right)}_{(c)}
   \underbrace{f\!\left(\overline{\boldsymbol{y}}_{n^{\prime}}^{(j)},\boldsymbol{b}_{n^{\prime}}^{(j)},m_{n^{\prime}}^{(j)} \,\big|\, \underline{\boldsymbol{y}}_{n^{\prime}}^{(j)} \,;\, \mathbf{u}_{n^{\prime}}\right)}_{(d)} \\
 & \times \underbrace{f\!\left(\underline{\boldsymbol{y}}_{n^{\prime}}^{(j)} \,\big|\, \overline{\boldsymbol{y}}_{n^{\prime}}^{(j)},\underline{\boldsymbol{y}}_{n^{\prime}}^{(j)}\right)}_{(e)}
   \underbrace{f\!\left(\overline{\tilde{\boldsymbol{y}}}_{n^{\prime}}^{(j)},\boldsymbol{c}_{n^{\prime}}^{(j)},\tilde{m}_{n^{\prime}}^{(j)} \,\big|\, \underline{\tilde{\boldsymbol{y}}}_{n^{\prime}}^{(j)} \,;\, \mathbf{u}_{n^{\prime}}\right)}_{(f)} \\
 & \times \underbrace{f\!\left(\boldsymbol{z}_{1:n}\mid \boldsymbol{y}_{0:n},\boldsymbol{b}_{1:n},\boldsymbol{m}_{1:n}\,;\,\mathbf{u}_{0:n}\right)}_{(g)}
   \underbrace{f\!\left(\tilde{\boldsymbol{z}}_{1:n}\mid \tilde{\boldsymbol{y}}_{0:n},\boldsymbol{c}_{1:n},\tilde{\boldsymbol{m}}_{1:n}\,;\,\mathbf{u}_{0:n}\right)}_{(h)}
   \label{eq:jointpdf2}
\end{aligned}
\end{equation}
\begin{figure*}[!b]
\centering
\vspace{-4.3mm}
\noindent\hrulefill
\vspace{-3pt} % 减少两个公式之间的间距
\begin{equation}\begin{aligned}
 f\big(&\boldsymbol{y}_{0:n},\tilde{\boldsymbol{y}}_{0:n},\boldsymbol{b}_{1:n},\boldsymbol{c}_{1:n},\boldsymbol{m}_{1:n},\tilde{\boldsymbol{m}}_{1:n}\mid\boldsymbol{z}_{1:n},\tilde{\boldsymbol{z}}_{1:n};\mathbf{u}_{0:n}\big) \\
 \propto& 
%%虚警和新feature的先验信息
\prod_{j^{\prime}=1}^{J}\underbrace{\left(\prod_{k=1}^{K_{0,\mathrm{bi}}^{(j')}}f\left(\boldsymbol{y}_{k,0}^{(j')};\mathbf{u}_{0}\right)\right)\left(\prod_{k'=1}^{K_{0,\mathrm{mo}}^{(j')}}f\left(\tilde{\boldsymbol{y}}_{k^{\prime},0}^{(j')};\mathbf{u}_{0}\right)\right)}_{\text{prior PDFs related to UAV and BS}}\left(\prod_{n^{\prime}=1}^{n}\underbrace{\left(\prod_{l=1}^{m_{n}^{(j')}}f_{\mathrm{fa}}({z}_{l,n^{\prime}}^{(j')})\right)\left(\prod_{l=1}^{\tilde{m}_{n}^{(j')}}f_{\mathrm{fa}}(\tilde{\boldsymbol{z}}_{l,n^{\prime}}^{(j')})\right)}_{\text{FA prior PDFs related to UAV and BS}}\right)\\
&\times\prod_{j=1}^{J}\prod_{n^{\prime}=1}^{n}\left\{\prod_{{k}=1}^{\underline{K}_{n^{\prime},\mathrm{bi}}^{(j)}}\underbrace{\underline{q}\left(\underline{\boldsymbol{x}}_{k,n^{\prime}}^{(j)},\underline{r}_{k,n^{\prime}}^{(j)}\mid\tilde{\boldsymbol{x}}_{k,n^{\prime}-1}^{(j)},\tilde{r}_{k,n^{\prime}-1}^{(j)}\right)}_{\text{state-transition factor corresponding to \emph{Stage 1}}}\prod_{l=1}^{m_{n^{\prime}}^{(j)}}\underbrace{h_{k}\left(\underline{\boldsymbol{x}}_{k,n^{\prime}}^{(j)},\underline{r}_{k,n^{\prime}}^{(j)},b_{l,n^{\prime}}^{(j)};{z}_{l,n^{\prime}}^{(j)},\mathbf{u}_{n^{\prime}}\right)}_{\text{likelihood factor for leagacy PMFs (bi)}} \right.
\\
& \times\prod_{k^{\prime}=1}^{m_{n^{\prime}}^{(j)}}\underbrace{\bar{q}\left(\overline{\boldsymbol{x}}_{k^{\prime},n^{\prime}}^{(j)},\overline{r}_{k^{\prime},n^{\prime}}^{(j)};\mathbf{u}_{n^{\prime}}\right)}_{\text{prior factor for new PMFs (bi)}}
\left.\underbrace{
g_{\underline{K}_{n,\mathrm{bi}}^{(j)}+k^{\prime}}\!\left(\overline{\boldsymbol{x}}_{k^{\prime},n'},\overline{r}_{k^{\prime},n'},b_{k^{\prime},n'}^{(j)};\boldsymbol{z}_{k^{\prime},n'},\mathbf{u}_{n'}\right)
\prod_{l=k^{\prime}+1}^{m_{n}^{(j)}}h_{\underline{K}_{n^{\prime},\mathrm{bi}}^{(j)}+k^{\prime}}\left(\overline{\boldsymbol{x}}_{k^{\prime},n^{\prime}}^{(j)},\overline{r}_{k^{\prime},n}^{(j)},b_{l,n^{\prime}}^{(j)};{z}_{l,n^{\prime}}^{(j)},\mathbf{u}_{n^{\prime}}\right)}_{\text{likelihood factor for new PMFs (bi) }}\right\}\\
&\times \left.\left\{\prod_{k=1}^{\underline{K}_{n',\mathrm{mo}}^{(j)}}\underbrace{\tilde{\underline{q}}\left(\underline{\tilde{\boldsymbol{x}}}_{k,n^{\prime}}^{(j)},\underline{\tilde{r}}_{k,n^{\prime}}^{(j)}\mid\boldsymbol{x}_{k,n^{\prime}}^{(j)},r_{k,n^{\prime}}^{(j)}\right)}_{\text{state-transition factor corresponding to \emph{Stage 2}}}\right.\right.
\prod_{l=1}^{\tilde{m}_{n^{\prime}}^{(j)}}\underbrace{\tilde{h}_{k}\left(\tilde{\underline{\boldsymbol{x}}}_{k,n^{\prime}}^{(j)},\tilde{\underline{r}}_{k,n^{\prime}}^{(j)},c_{l,n^{\prime}}^{(j)};\tilde{\boldsymbol{z}}_{l,n^{\prime}}^{(j)},\mathbf{u}_{n^{\prime}}\right)}_{\text{likelihood factor for leagacy PMFs (mo)}}\\
&\times\prod_{k^{\prime}=1}^{\tilde{m}_{n^{\prime}}^{(j)}}\underbrace{\bar{\tilde{q}}\left(\overline{\tilde{\boldsymbol{x}}}_{k^{\prime},n^{\prime}}^{(j)},\overline{\tilde{r}}_{k^{\prime},n^{\prime}}^{(j)};\mathbf{u}_{n^{\prime}}\right)}_{\text{prior factor for new PMFs (mo)}}
\underbrace{\tilde{g}_{\underline{K}_{n^{\prime},\mathrm{mo}}^{(j)}+k^{\prime}}\left(\overline{\tilde{\boldsymbol{x}}}_{k^{\prime},n^{\prime}}^{(j)},\overline{\tilde{r}}_{k^{\prime},n}^{(j)},c_{k,n^{\prime}}^{(j)};\tilde{\boldsymbol{z}}_{k,n^{\prime}}^{(j)},\mathbf{u}_{n^{\prime}}\right)
\prod_{l=k^{\prime}+1}^{\tilde{m}_{n}^{(j)}}\tilde{h}_{\underline{K}_{n^{\prime},\mathrm{mo}}^{(j)}+k}\left(\overline{\tilde{\boldsymbol{x}}}_{k^{\prime},n^{\prime}}^{(j)},\overline{\tilde{r}}_{k^{\prime},n}^{(j)},c_{l,n^{\prime}}^{(j)};\tilde{\boldsymbol{z}}_{l,n^{\prime}}^{(j)},\mathbf{u}_{n^{\prime}}\right)}_{\text{likelihood factor for leagacy PMFs (mo)}}
\label{eq:F2}
\end{aligned}\end{equation}
\noindent\hrulefill
\end{figure*}
%%下面的得再修改
\vspace{-14pt}
\par\noindent where (a) and (b) denote the prior factor for the initial PMFs for both link; if no prior is available, this term can be omitted.
(c) and (e) are the state-transition PDF specified in Sec. III.
(d) is the data-association factor between measurements and PMFs for the bistatic link, and (f) is its counterpart for the monostatic link. (g) and (h) encode the measurement likelihoods for the bistatic and monostatic links, respectively.
\begin{figure}[htbp]
    \centering
    \begin{subfigure}{0.25\textwidth}
        \centering
        \includegraphics[width=\textwidth]{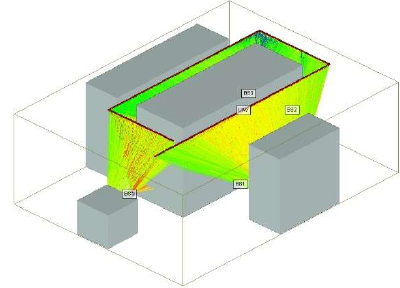}
        \caption{}
        \label{fig:sub1}
    \end{subfigure}
    \hfill
    \begin{subfigure}{0.22\textwidth}
        \centering
        \includegraphics[width=\textwidth]{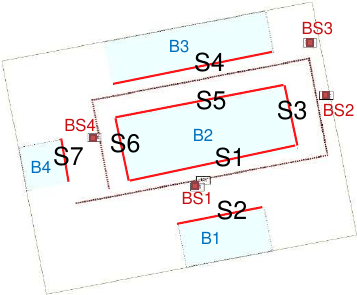}
        \caption{}
        \label{fig:sub2}
    \end{subfigure}
    \caption{Scenario configuration. (a) 3D view, (b) 2D view.}
    \label{fig:scenario}
    \vspace{-10pt}
\end{figure}

Using common assumptions \cite{Leitinger2019BP_SLAM},\,\,\cite{bar-shalom2011tracking}, the joint
posterior PDF in (\ref{eq:jointpdf2}) can be factorized as (\ref{eq:F2}), where $\underline{q}\left(\cdot\right)$, $\tilde{\underline{q}}\left(\cdot\right)$, $h_{k}\left(\cdot\right)$, $\tilde{h}_{k}\left(\cdot\right)$,  $\overline{q}\left(\cdot\right)$, $\overline{\tilde{q}}\left(\cdot\right)$, ${g}_{\underline{K}_{n'}^{(j)}+k'}\left(\cdot\right)$, $\tilde{g}_{\underline{K}_{n'}^{(j)}+k'}\left(\cdot\right)$, ${h}_{\underline{K}_{n'}^{(j)}+k'}\left(\cdot\right)$ and $\tilde{h}_{\underline{K}_{n'}^{(j)}+k'}\left(\cdot\right)$   follow the same structure as introduced in Sec. IV-B.
%% 方案2视角交替的因子分解如下：

\begin{comment}
    \subsection{Inference Strategy on Factor Graphs}
This section will describe how the proposed Bayesian frameworks are implemented through inference on factor graphs. Since our factor graph in Fig. 2 has cycles, we have to decide on a specific order of message computation \textcolor{blue}{cite}. We
choose this order according to the following rules: (i) messages are not sent backward in time \textcolor{blue}{cite}; and (ii) at each time step messages are computed and processed in parallel. With these rules, the generic message passing equations of the SPA \textcolor{blue}{cite} yield the following operations at each time step. The corresponding messages are shown in Fig. 2.
\end{comment}

\subsection*{D. Inference Strategy and Computational Complexity and Parallelization Analysis}
The proposed factor graph can be efficiently inferred via the sum-product algorithm (SPA)~\cite{9585528}. 
The per-epoch complexities are $\mathcal{O}(K_{\mathrm{bi}}M_{\mathrm{bi}}+K_{\mathrm{bi}}M_{\mathrm{mo}})$ for Scheme~I and 
$\mathcal{O}(K_{\mathrm{bi}}M_{\mathrm{bi}})+\mathcal{O}(K_{\mathrm{mo}}M_{\mathrm{mo}})$ for Scheme~II. 
When $K_{\mathrm{bi}}\approx K_{\mathrm{mo}}$ and $M_{\mathrm{bi}}\approx M_{\mathrm{mo}}$, the two schemes have comparable asymptotic complexity but differ in parallelization and latency.
\textcolor{black}{
Here, $M_{\mathrm{bi}}$ and $M_{\mathrm{mo}}$ denote the numbers of detected MPCs per epoch on the bistatic and monostatic links, respectively, so denser multipath mainly increases $M$. 
The cost scales bilinearly with the number of maintained active PMFs and measurements (i.e., $K \times M$). 
In practice, the effective number of resolvable dominant MPCs is limited by array/delay resolution and detection thresholds. If needed, preprocessing techniques such as measurement clustering or censoring can further reduce the effective measurement load (see~\cite{9585528}, Sec.~IV-D).}
As illustrated in Fig.~\ref{fig:factorPic}, Scheme~I uses a unified graph and performs parallel message updates across link-specific likelihood factors within each iteration. 
Scheme~II runs the two link-specific modules sequentially and propagates the inferred marginals across links via the state-transition factors, which introduces inter-module synchronization and lengthens the per-epoch critical path. Consequently, Scheme~I typically yields lower per-epoch latency and is preferable for latency-sensitive applications. \textcolor{black}{With pruning, the active PMF counts $K_{\mathrm{bi}}$ and $K_{\mathrm{mo}}$ remain bounded, so the per-epoch runtime stays stable over long observation windows.}

\begin{figure*}
    \centering
    \includegraphics[width=1\linewidth]{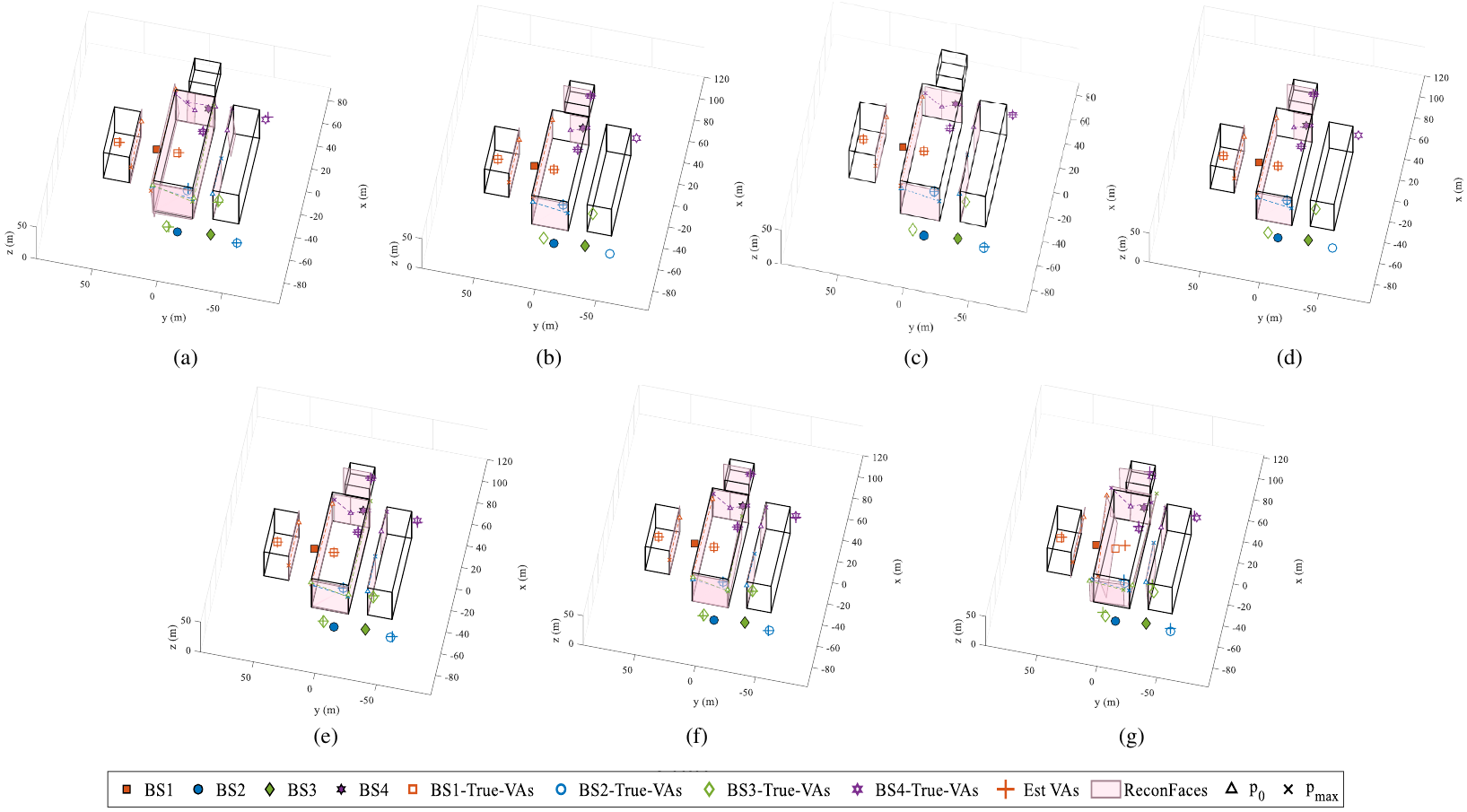}
    \caption{Comparison of environment map reconstruction results for different approaches. (a) Bistatic-only sensing \cite{Wielandner2023NonIdealSurfaces} , (b) monostatic-only sensing, (c) proposed fusion 1, (d) proposed fusion 2, (e) proposed fusion 3, (f) proposed fusion 4, (g) fusion approach in \cite{Yang2022JSACHybridActivePassive}.}
        \label{fig:map}
        \vspace{-5pt}
\end{figure*}

\section{Experimental Evaluation}
This section evaluates the performance of the proposed  approach using synthetic RF data generated by Wireless InSite, a widely-used commercial ray-tracing software,  under realistic rough-surface propagation conditions. We benchmark our method against three reference approaches: monostatic-only sensing, bistatic-only sensing \cite{Wielandner2023NonIdealSurfaces}, and a fusion variant that integrates both sensing modes but neglects diffuse scattering effects.
The comparison focuses on the accuracy and completeness of facade localization, which are critical for reconstructing the propagation environment in ISAC-enabled low-altitude networks.

\subsection{Experimental Setup}

\textcolor{black}{
To validate the proposed fusion approach, we construct a representative urban scenario in \emph{Wireless InSite}, as illustrated in Fig.~3. The environment is partitioned into three \emph{spatial facade blocks} based on the top-view adjacency in Fig.~3(b): {Block~I} $(S_1,S_2)$, {Block~II} $(S_3,S_4,S_5)$, and {Block~III} $(S_6,S_7)$. Fig.~3(a) shows the corresponding 3D view, and Fig.~3(b) depicts the top view with the deployed BSs and the UAV sampling locations.}

\textcolor{black}{
To capture link-dependent visibility heterogeneity, we further categorize facades into \emph{observability classes} according to the available measurement types. {Class~A} (\emph{dual-link, fully observable}) facades are visible to both monostatic and bistatic links, i.e., $\{S_1,S_2,S_3,S_6\}$. {Class~B} (\emph{bistatic-only, partially observable}) facades lose monostatic/backscattered paths due to geometric occlusion and are observable only via bistatic links, i.e., $\{S_4,S_5\}$. {Class~C} (\emph{monostatic-only, partially observable}) facade $S_7$ lacks viable bistatic paths and is observable only through the monostatic link. As a result, {Block~I} is homogeneous in observability, whereas {Blocks~II--III} are heterogeneous, containing a mixture of fully and partially observable facades.}

\textcolor{black}{
The scene contains four concrete buildings $B_1$--$B_4$ centered at $(0,48,20)$~m, $(0,0,24)$~m, $(0,-40,22.5)$~m, and $(75,0,8)$~m, with dimensions $40{\times}20{\times}40$~m, $80{\times}30{\times}48$~m, $75{\times}20{\times}45$~m, and $20{\times}20{\times}16$~m, respectively; material parameters follow ITU-R P.2040-4~\cite{ITUR_P2040_4_2025}. Four monostatic BSs are placed at $(8,23,8)$~m, $(-58,-7,8)$~m, $(-55,-32,8)$~m, and $(50,-8,8)$~m, operating at $30$~GHz with $30$~dBm transmit power. The UAV flies at $10$~m/s and samples every $0.1$~s for $T=305$ time steps. To emulate snapshot-to-snapshot fluctuations in monostatic backscatter, we sample a small neighborhood around each BS by adding $304$ auxiliary receiver locations in addition to the nominal BS position, and treat the resulting backscatter snapshots as time-indexed observations aligned with the $T$ UAV steps.}

\textcolor{black}{Given the above deployment and sampling protocol, \emph{Wireless InSite} is used as a ray-tracing (RT) engine to generate geometry-consistent propagation truth for both bistatic (BS--UAV) and monostatic (BS-backscatter) links. Since RT outputs represent the ideal ground truth, we convert them into measurement-level pseudo observations by adding sensing non-idealities. Specifically, RT delays are mapped to bistatic pseudo-range measurements, and RT scattering-point locations are treated as nominal monostatic pseudo-position measurements. To simulate realistic measurement conditions, we inject (i) additive estimation noise, (ii) i.i.d.\ Poisson clutter (false alarms).}

\textcolor{black}{
\emph{Measurement uncertainty (consistent with Sec.~II-B).}
For the bistatic link, the delay-estimation error is modeled as
$E^{\mathrm{bi},(j)}_{m,n}\sim\mathcal N(0,\sigma_\tau^2)$, which maps to the pseudo-range domain as
$\sigma_z^2=\mathrm c^2\sigma_\tau^2$ in \eqref{eq:ZcondBiGaussian}; in the simulations we set the pseudo-range standard deviation to $\sigma_z=0.5$~m.
For the monostatic link, the pseudo-position is obtained via the converted-measurement mapping $\mathbf g^{(j)}(\cdot)$ in \eqref{eq:gDefMo}--\eqref{eq:pseudoposDefMo}.
Under a first-order approximation, the equivalent pseudo-position-domain error is modeled as
$\boldsymbol{E}^{\mathrm{mo},(j)}_{m,n}\sim\mathcal N(\mathbf 0,\mathbf R^{(j)}_{m,n})$ in \eqref{eq:measureModelMo}, where
$\mathbf R^{(j)}_{m,n}$ can be obtained from the front-end delay/angle covariance $\boldsymbol{\Sigma}^{(j)}_{\eta,m,n}$ via Jacobian/Taylor propagation, i.e.,
$\mathbf R^{(j)}_{m,n}\approx \mathbf J^{(j)}_{g,m,n}\boldsymbol{\Sigma}^{(j)}_{\eta,m,n}\big(\mathbf J^{(j)}_{g,m,n}\big)^{\mathsf T}$.
As a transparent baseline, we set $\mathbf R^{(j)}_{m,n}=\sigma_{\mathrm{BS}}^{2}\mathbf I_{3}$ with $\sigma_{\mathrm{BS}}=0.1~\text{m}$, i.e., an isotropic (diagonal) \emph{per-axis} pseudo-position covariance. This choice is used only in the experimental setup to avoid imposing a specific, estimator-dependent correlation structure; it is \emph{not} a limitation of the proposed inference, since the monostatic likelihood in Sec.~III-C accepts a general anisotropic, range-dependent $\mathbf R^{(j)}_{m,n}$ and propagates it to the projection-residual covariance accordingly.
In the RT-based simulation, the scattering-point locations provided by \emph{Wireless InSite} are treated as the nominal pseudo-positions (i.e., $\mathbf g^{(j)}(\boldsymbol{\eta}^{\mathrm{mo},(j)}_{m,n})$), and the converted-measurement uncertainty is injected directly in the pseudo-position domain via $\boldsymbol{E}^{\mathrm{mo},(j)}_{m,n}\sim\mathcal N(\mathbf 0,\mathbf R^{(j)}_{m,n})$.
Finally, the adopted uncertainty magnitudes $\sigma_z=0.5$~m and $\sigma_{\mathrm{BS}}=0.1$~m fall into a representative sub-meter/decimeter regime and are used as order-of-magnitude anchors consistent with, e.g., TS~22.125, TS~22.186, and the 5G Americas ISAC white paper, rather than hardware-specific error budgets~\cite{3gpp_ts_22_125_v18_1_0,3gpp_ts_22_186_v19_0_0,5gamericas_isac_2025}.}

\textcolor{black}{
\emph{Clutter model.}
We adopt a standard i.i.d.\ clutter model to emulate practical detection artifacts. Specifically, the numbers of false-alarm measurements are Poisson distributed with mean $\mu_{\mathrm{fa}}=1$ for bistatic links and $\tilde{\mu}_{\mathrm{fa}}=1$ for monostatic links. This setting represents a moderate false-alarm load per epoch, consistent with thresholded MPC detection where only a small number of spurious peaks survive after screening.  Conditioned on their counts, bistatic false alarms are drawn uniformly over the pseudo-range interval $[0,500]$~m, while monostatic false alarms are drawn uniformly over $[-150,150]\times[-150,150]\times[-150,150]$~m$^3$ in 3D space.}

\textcolor{black}{
\emph{Birth/survival and cross-link settings.}
New PMF/VA hypotheses are instantiated in a measurement-driven manner using the birth prior $f_{\mathrm n}(\boldsymbol{x}^{(j)}_{k,n};\mathbf u_n)$, which is set uniform over the region of interest (ROI), modeled as a sphere of radius $100$~m centered at the scenario in Fig.~3. The corresponding mean birth intensities are set to $\mu_{\mathrm{n},n}^{(j)}=\tilde{\mu}_{\mathrm{n},n}^{(j)}=0.01$, reflecting sparse hypothesis-instantiation events per time step within this local ROI.
The survival probability is set to $P_{\mathrm s}=0.99$ to reflect the quasi-static nature of building facades (and thus VA hypotheses) over the short inter-epoch interval, while allowing occasional termination under prolonged missed detections or association ambiguity. The expected numbers of detected MPCs per facade, $\mu_{\mathrm m}$ and $\tilde{\mu}_{\mathrm m}$, are determined from noise-free RT data via power thresholding and statistical averaging, yielding $\mu_{\mathrm m}=\tilde{\mu}_{\mathrm m}=4$.
For Scheme~II, we introduce the cross-link parameters $P_{\mathrm c}$ and $P_{\mathrm b}$ to accommodate a more general view-dependent cross-link visibility/transition model (Sec.~IV). In this paper, we adopt the deterministic setting $P_{\mathrm c}=1$ and $P_{\mathrm b}=0$, consistent with our sequential set-update protocol: hypotheses are carried as candidates across links, while new hypotheses are introduced explicitly by set augmentation with the current measurement set, and view-dependent intermittency is handled through the clutter model and the link-specific likelihoods (Sec.~III-C), rather than through probabilistic cross-link birth/persistence. More general choices ($P_{\mathrm c}<1$ and/or $P_{\mathrm b}>0$) can be used to explicitly model strongly intermittent cross-link visibility without changing the proposed fusion framework.
In larger or denser environments, the clutter/birth parameters can be calibrated from site measurements or inferred online using intensity/PHD-type estimators \cite{horridge2011using}.}

\textcolor{black}{
\emph{Inference configuration.}
We run 2 iterations of sum-product message passing per epoch, using a particle-based representation for continuous PMF states with 20{,}000 particles.
A PMF is declared to exist if $\Pr(r^{(j)}_{k,n}=1)>P_{\mathrm{th}}$ (we set $P_{\mathrm{th}}=0.5$).
To keep the number of active PMFs bounded, we apply pruning with threshold $P_{\mathrm{prun}}=10^{-3}$ by discarding PMFs with $\Pr(r^{(j)}_{k,n}=1)<P_{\mathrm{prun}}$, thereby improving computational efficiency.}

\textcolor{black}{
\emph{Remarks on practical deployments.}
Real deployments may introduce additional impairments (e.g., synchronization offsets, phase noise, calibration errors, UAV attitude jitter, and environmental dynamics), which can lead to biased/non-Gaussian errors and spatially non-uniform clutter. Such effects can be accommodated within the proposed Bayesian framework by refining the measurement likelihood and clutter/cardinality models (e.g., conservative covariance inflation, heavier-tailed likelihoods, or learned clutter intensities), without altering the underlying fusion mechanism.}

%\footnote{It is worth noting that per-epoch measurement counts on each façade may deviate from a Poisson law in practice owing to material heterogeneity, UAV motion and attitude instability, sensing geometry, and other site specific factors. Nevertheless, the proposed Bayesian fusion framework remains applicable by replacing the Poisson cardinality model with a more suitable count distribution (e.g., binomial, negative binomial, or an inhomogeneous Poisson process) and by updating the associated likelihood and message-passing terms accordingly.}
\subsection{Visualization of Environment Map Completeness and Accuracy}
This subsection provides a qualitative comparison of the reconstructed environment maps to assess map completeness and geometric fidelity; quantitative errors are reported in the next subsection.
Fig.~\ref{fig:map} shows the maps obtained by different approaches.

\emph{Visualization procedure.}
To suppress snapshot-level fluctuations, we apply a short temporal smoothing for visualization: each VA estimate at its last detection time is averaged with the preceding five time steps.
Using the smoothed VA positions, each facade is recovered as the perpendicular-bisector plane of the VA--BS segment; for rendering, we assign a finite extent based on the visible UAV trajectory (width estimated from the maximal horizontal spread of VA-induced intersection points, and height set to a typical urban-building value).

\emph{Qualitative comparison.}
Single-link methods exhibit an inherent coverage--accuracy trade-off.
Bistatic-only sensing~\cite{Wielandner2023NonIdealSurfaces} (Fig.~\ref{fig:map}(a)) achieves broader coverage (6/7 facades) but suffers from larger VA localization errors due to noisier bistatic measurements.
Monostatic-only sensing (Fig.~\ref{fig:map}(b)) yields higher localization accuracy but is visibility-limited (5/7 facades), failing to detect the canyon-occluded facades S4--S5.
The baseline fusion method~\cite{Yang2022JSACHybridActivePassive} (Fig.~\ref{fig:map}(g)) uses monostatic measurements mainly for initialization and assumes purely specular one-to-one association; without explicit diffuse-scattering modeling, it becomes brittle in rough-surface environments and produces the largest mismatch to the ground truth.
\begin{figure} \centering \includegraphics[width=0.75\linewidth]{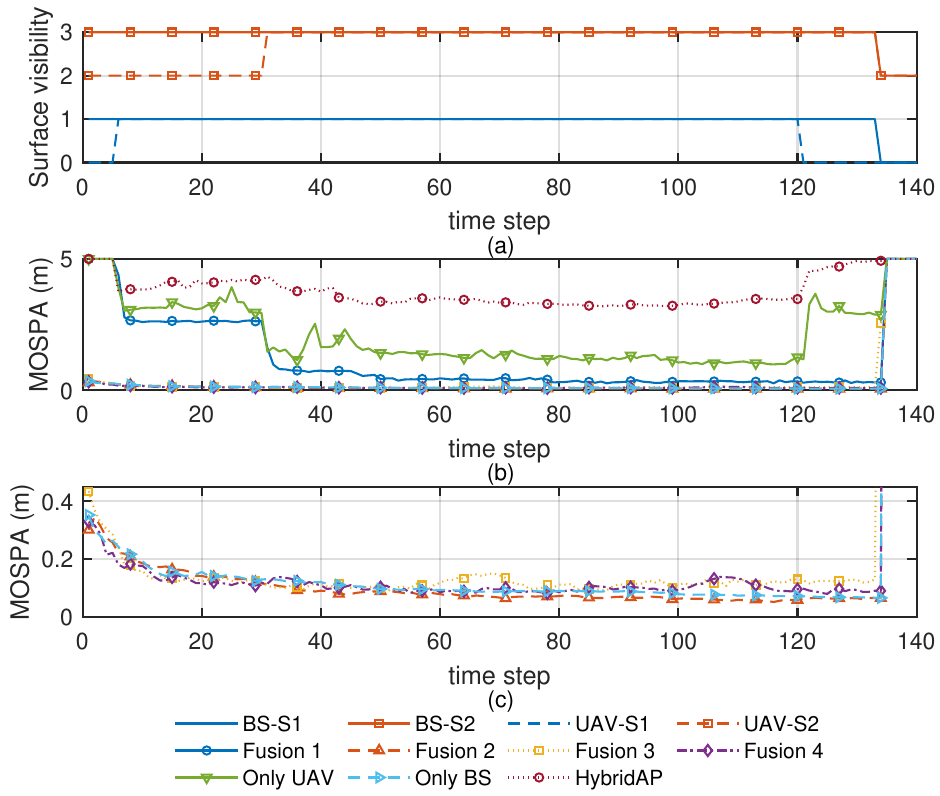} \caption{Mapping performance of BS1 for facades visible on both bistatic and monostatic links.
(a) Ground-truth facade visibility versus time. S1 and S2 are encoded as ${0,1}$ and ${2,3}$, respectively, where 0/2 denote non-visible states and 1/3 denote visible states. Solid lines denote BS backscatter observations, and dashed lines denote BS--UAV bistatic observations. S1 is shown without markers, and S2 with square markers.
(b) MOSPA error versus time for the four proposed fusion approaches, the bistatic-only scheme (Only UAV), the monostatic-only scheme (Only BS), and the baseline HybridAP method \cite{Yang2022JSACHybridActivePassive}. Lower MOSPA indicates higher localization accuracy.
(c) Enlarged view of (b) for a clearer comparison in the low-error region.} \label{fig:mospa1} \vspace{-10pt} \end{figure}

Compared with these baselines, the proposed fusion schemes (Figs.~\ref{fig:map}(c)--(f)) improve both robustness and map quality.
Scheme~I (Fusion~1--2) enhances estimation by incorporating auxiliary measurements from the other link, while its map coverage remains determined by the dominant link.
Scheme~II (Fusion~3--4) performs sequential cross-link propagation through the state-transition factor and achieves the most complete reconstruction, successfully recovering all observable facades (7/7) while maintaining localization accuracy comparable to Scheme~I.

\subsection{Performance of the Proposed Fusion Mechanism}
In this subsection, we evaluate mapping performance using MOSPA, i.e., the OSPA-based error averaged over 100 Monte-Carlo trials. We set the OSPA cut-off (cardinality-penalty) parameter to \(5\), so missed/false facades incur a bounded penalty up to this value. 
We evaluate four BS configurations (BS~1--BS~4) that cover distinct link-visibility/observability conditions in Fig.~3, thereby probing robustness under heterogeneous information availability.

Fig.~\ref{fig:mospa1} quantifies the mapping performance under the BS1 configuration. First, the visibility timeline in Fig.~\ref{fig:mospa1}(a) provides the causal context for interpreting MOSPA: the monostatic link observes both facades from the beginning, whereas bistatic visibility emerges later (at time steps 7 and 31 for S1 and S2) and becomes unavailable for S1 after time step 120. These visibility transitions correspond to the turning points and regime changes in the MOSPA trajectories.
\begin{figure}
        \centering
        \includegraphics[width=0.75\linewidth]{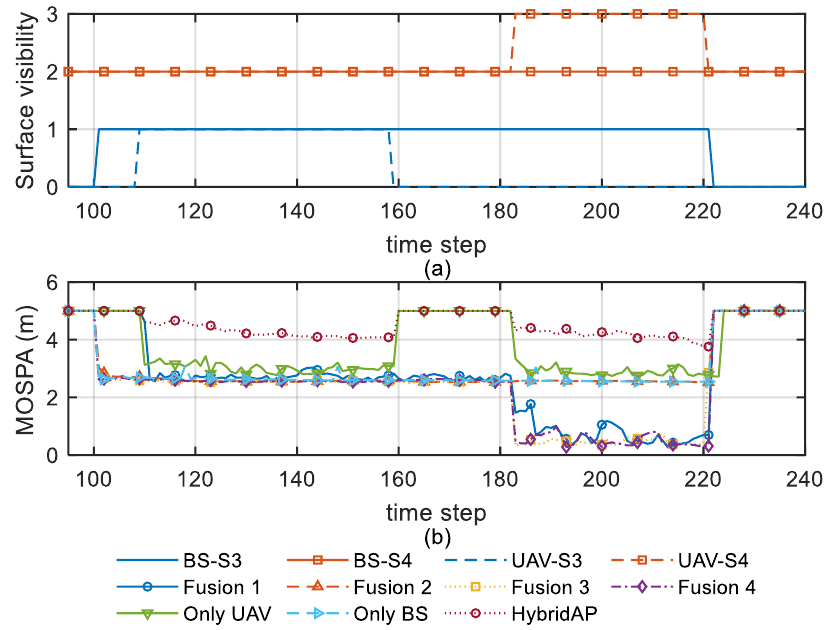}
        \caption{Mapping performance of BS2 for mutually visible facades S3 and S4. Notations and conventions follow Fig.~\ref{fig:mospa1}.}
                \label{fig:mospa2}
                \vspace{-3pt}
    \end{figure}
 \begin{figure}
        \centering
        \includegraphics[width=0.8\linewidth]{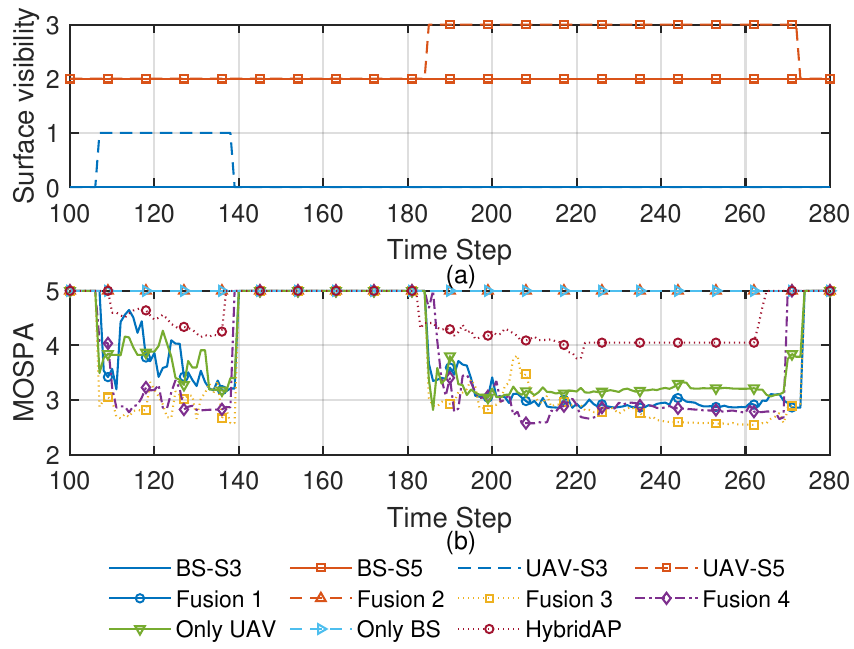}
        \caption{Mapping performance of BS3 for mutually visible facades S3 and S5. Notations and conventions follow Fig.~\ref{fig:mospa1}.}
        \label{fig:mospa3}
        \vspace{-3pt}
 \end{figure}
The MOSPA comparison in Fig.~5(b) highlights the coupled importance of diffuse-scattering modeling and data-level fusion. The baseline fusion method in~\cite{Yang2022JSACHybridActivePassive}, despite combining two sensing modes, performs the worst in this diffuse-scattering setting, even underperforming the bistatic-only reference \cite{Wielandner2023NonIdealSurfaces} that explicitly accounts for diffuse reflections. This indicates that unmodeled diffuse scattering can dominate the overall error budget and that dual-link fusion alone cannot compensate for such model mismatch. Meanwhile, although the bistatic-only method \cite{Wielandner2023NonIdealSurfaces} improves upon~\cite{Yang2022JSACHybridActivePassive}, its accuracy remains limited by sparse bistatic observations, substantial UAV measurement noise, and weak time-delay information, suggesting that diffuse modeling must be coupled with effective dual-link fusion to fully realize its benefit.
Considering the map-completeness results, Fusion~2 and  monostatic-only are intrinsically constrained by the monostatic field-of-view and thus provide limited coverage; hence, the following discussion focuses on strategies that preserve extensive map coverage. The proposed Scheme~II (Fusion~3--4) achieve the best performance: once both facades are observable, they attain the lowest MOSPA with stable convergence (see Fig.~5(c)), yielding an approximately 90\% error reduction relative to the bistatic-only method. Fusion~3 and Fusion~4 behave similarly because they differ only in the direction of cross-link information propagation, allowing implementation to be selected based on the availability of initial priors.
Although Scheme~I (Fusion~1) is not optimal due to bistatic-dominant noise, it remains practically valuable. Its MOSPA exhibits pronounced drops aligned with the onset of bistatic visibility (time steps 7 and 31), and by time step 120 it achieves a 67\% error reduction relative to bistatic-only. More importantly, once S1 becomes bistatic-invisible after time step 120, bistatic-only deteriorates due to the lack of further feature updates, whereas Fusion~1 continues refining the estimate using monostatic backscatter, leading to sustained error reduction; this robustness is further evidenced under the BS2/BS3 configurations in Figs.~\ref{fig:mospa2}--\ref{fig:mospa3}.
 \begin{figure}
        \centering
        \includegraphics[width=0.85\linewidth]{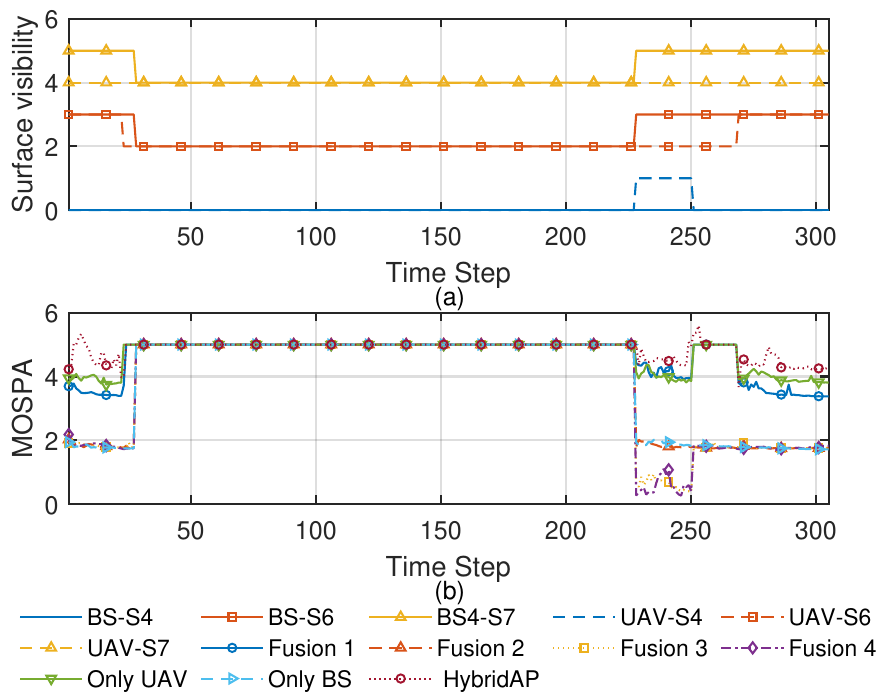}
        \caption{Mapping performance of BS4 for mutually visible facades S4, S6, and S7. Notations and conventions follow Fig.~\ref{fig:mospa1}.}
        \label{fig:mospa4}
        \vspace{-10pt}
 \end{figure}
 
Fig.~\ref{fig:mospa2} further evaluates the link-limited visibility cases, where some facades are observable only via bistatic NLoS multipath while monostatic backscatter is unavailable, leading to a pronounced imbalance between the two sensing modes. Overall, both Scheme~I (Fusion~1) and Scheme~II (Fusion~3--4) achieve consistently lower MOSPA than the bistatic-only baseline, demonstrating clear robustness under partial information loss. A representative example occurs during time steps 160--180: as shown in Fig.~\ref{fig:mospa2}(a), facade S3 becomes bistatic-invisible in this interval, so the bistatic-only method cannot further update the corresponding feature and its error increases abruptly. In contrast, although Fusion~1 is bistatic-dominated, S3 has already been instantiated before the outage and remains observable on the monostatic link; hence Fusion~1 continues refining S3 using monostatic backscatter and maintains stable (or even improving) localization accuracy. During 180--220, the bistatic link starts observing a new facade S4, which drives down the bistatic-only error; nevertheless, Fusion~1 remains substantially better because it simultaneously sustains continuous monostatic updates for S3. Scheme~II (Fusion~3--4) consistently delivers the lowest MOSPA by more effectively exploiting the complementary updates available across the two modes whenever they exist.

Fig.~\ref{fig:mospa3} considers the BS3 configuration, where monostatic backscatter is essentially absent for all observable facades (Fig.~\ref{fig:mospa3}(a)), so the system effectively reduces to bistatic-only sensing. In this limit, the proposed fusion strategies naturally converge toward the bistatic-only performance in Fig.~\ref{fig:mospa3}(b), confirming that additional fusion gains require genuinely complementary cross-link information. Importantly, this case also demonstrates a desirable graceful-degradation property: when one sensing mode becomes uninformative, the proposed framework does not exhibit instability, but smoothly reduces to the informative single-link bound, whereas the  baseline~\cite{Yang2022JSACHybridActivePassive} remains noticeably inferior under such severe information deficiency.

Fig.~\ref{fig:mospa4} evaluates the BS4 configuration, where some facades are observable only through monostatic backscatter, remaining undetectable on the bistatic link. Specifically, S7 is bistatic-invisible throughout the sequence and can only be observed via monostatic backscatter. Interestingly, the same scenario also exhibits a complementary link-exclusive pattern: S4 remains monostatic-invisible and becomes bistatic-visible only over a short interval (approximately 230-250). Such mode-split visibility implies an inherent limitation of single-link mapping, since each method is confined to the facade set observable by its own sensing mode and thus cannot recover both S7 and S4. Scheme~II addresses this by sequential cross-link propagation through the state-transition factor, enabling the fusion process to exploit the union of observations from the two sensing modes. Consequently, Fusion~3/4 exhibits the sharpest error drop and attains the lowest MOSPA during 230-250, demonstrating Scheme~II's ability to exploit complementary visibility across the two sensing modes through cross-link propagation.

Overall, the two schemes offer complementary trade-offs: Scheme~I favors link-dominant updating with auxiliary fusion, whereas Scheme~II favors cross-link propagation to improve map completeness under heterogeneous link visibility.
\subsection*{\textcolor{black}{D. Runtime comparison.}}
\textcolor{black}{
To empirically validate the latency advantage of Scheme~I discussed in
Sec.~IV-D, we report wall-clock runtime measurements per sensing epoch
under varying measurement counts $M$ and particle numbers $N_p$. The runtime for each epoch is computed by dividing the total runtime over
all $T$ epochs by $T$.
The experiments are implemented in MATLAB~R2023b on a workstation
equipped with an Intel Core i9-14900HX CPU and 32~GB RAM.
Here $M$ denotes the number of detected MPC measurements associated
with each facade on each sensing link (bistatic and monostatic),
obtained from the ray-tracing outputs after power-threshold filtering.
$N_p$ denotes the number of particles used to represent the continuous
PMF states.
Table~\ref{tab:runtime_scaling} summarizes the measured per-epoch runtimes
(in seconds) for Scheme~I and Scheme~II. Each entry reports the runtime of
\textit{Scheme~I / Scheme~II}. The speedup is defined as
\[
\text{Speedup} =
\frac{\text{runtime of Scheme II}}
{\text{runtime of Scheme I}},
\]
and is reported for $N_p=20{,}000$, which corresponds to the particle
configuration used in the main experiments.
From the table, Scheme~I consistently achieves lower runtime than
Scheme~II, with an observed speedup ranging from $1.86$ to $1.90\times$
across the tested scenarios, supporting its suitability for
latency-sensitive real-time applications. In contrast, Scheme~II achieves higher map completeness, making it preferable for scenarios where coverage and reliability are prioritized.}
\textcolor{black}{
\begin{table}[!t]
\caption{\textcolor{black}{Wall-clock latency (s/epoch) versus measurement count $M$ and particle number $N_p$ (per BS). 
Each entry reports \textit{Scheme~I / Scheme~II}. The speedup is computed as $\text{II}/\text{I}$ at $N_p=20{,}000$.}}
\label{tab:runtime_scaling}
\centering
\footnotesize
\setlength{\tabcolsep}{5pt}
\renewcommand{\arraystretch}{1.08}
\color{black}
\begin{tabular}{c|ccc|c}
\hline
\multirow{2}{*}{$M$} & \multicolumn{3}{c|}{$N_p$ (particles)} & \multirow{2}{*}{Speedup} \\
\cline{2-4}
& $5{,}000$ & $10{,}000$ & $20{,}000$ &  \\
\hline
$5$  & \textit{0.057 / 0.107} & \textit{0.111 / 0.206} & \textit{0.222 / 0.417} & \textit{1.881} \\
$10$ & \textit{0.103 / 0.191} & \textit{0.207 / 0.373} & \textit{0.408 / 0.762} & \textit{1.867} \\
$20$ & \textit{0.201 / 0.368} & \textit{0.401 / 0.756} & \textit{0.781 / 1.481} & \textit{1.897} \\
\hline
\end{tabular}
\end{table}
}

\section{Conclusion}
This paper developed a Bayesian multipath-based environment mapping framework for ISAC that jointly exploits bistatic NLoS multipath and monostatic backscatter under non-ideal (diffuse-scattering) facade propagation. By introducing a unified surface-based geometric model linking the two sensing modes, we enabled principled data-level association and fusion within a factor-graph formulation, and proposed two complementary inference schemes with different information-flow schedules. Simulations with synthetic RF data show that the proposed fusion consistently improves mapping accuracy, robustness, and convergence compared with single-link baselines and an existing hybrid method. In particular, Scheme~II achieves the highest map completeness when link visibility is heterogeneous and partially observable across modes, while maintaining high localization precision; Scheme~I provides a low-latency and parallelizable alternative that remains robust to intermittent link visibility loss. Future work will focus on over-the-air measurement validation and extensions to joint self-localization and multi-BS cooperative mapping.

%\vfill

%\begingroup
\footnotesize % 将参考文献字体设置为小号
%\scriptsize
%\bibliographystyle{unsrt}
\bibliographystyle{IEEEtran}
\bibliography{ref}

\begin{thebibliography}{10}
\providecommand{\url}[1]{#1}
\csname url@samestyle\endcsname
\providecommand{\newblock}{\relax}
\providecommand{\bibinfo}[2]{#2}
\providecommand{\BIBentrySTDinterwordspacing}{\spaceskip=0pt\relax}
\providecommand{\BIBentryALTinterwordstretchfactor}{4}
\providecommand{\BIBentryALTinterwordspacing}{\spaceskip=\fontdimen2\font plus
\BIBentryALTinterwordstretchfactor\fontdimen3\font minus \fontdimen4\font\relax}
\providecommand{\BIBforeignlanguage}[2]{{%
\expandafter\ifx\csname l@#1\endcsname\relax
\typeout{** WARNING: IEEEtran.bst: No hyphenation pattern has been}%
\typeout{** loaded for the language `#1'. Using the pattern for}%
\typeout{** the default language instead.}%
\else
\language=\csname l@#1\endcsname
\fi
#2}}
\providecommand{\BIBdecl}{\relax}
\BIBdecl

\bibitem{10955337}
Y.~Jiang \emph{et~al.}, ``Integrated sensing and communication for low altitude economy: Opportunities and challenges,'' \emph{IEEE Commun. Mag.}, early access, 2025, doi:{\color{black}\href{http://dx.doi.org/ 10.1109/MCOM.001.2400685}{ 10.1109/MCOM.001.2400685}}.

\bibitem{LowAltitudeISACWhitePaper2024}
\BIBentryALTinterwordspacing
{China Telecom} \emph{et~al.}, ``The low-altitude network by integrated sensing and communication ({ISAC}),'' Industry Consortium, White Paper, Feb. 2024. [Online]. Available: \url{https://www.zte.com.cn/content/dam/zte-site/res-www-zte-com-cn/mediares/zte/%E6%97%A0%E7%BA%BF%E6%8E%A5%E5%85%A5/%E7%99%BD%E7%9A%AE%E4%B9%A6/Low_altitude_network_by_ISAC.pdf}
\BIBentrySTDinterwordspacing

\bibitem{Mu2023UAVmeetsISAC}
J.~Mu \emph{et~al.}, ``{UAV} meets integrated sensing and communication: Challenges and future directions,'' \emph{IEEE Commun. Mag.}, vol.~61, no.~5, pp. 62--67, Jan. 2023.

\bibitem{11072035}
X.~Ye \emph{et~al.}, ``Integrated sensing and communications for low-altitude economy: A deep reinforcement learning approach,'' \emph{IEEE Trans. Wireless Commun.}, early access, 2025, doi:{\color{black}\href{http://dx.doi.org/10.1109/TWC.2025.3583950}{10.1109/TWC.2025.3583950}}.

\bibitem{Zhang2025FourSteps6GAIComMag}
J.~Zhang \emph{et~al.}, ``Four steps toward {6G} {AI}-enabled air interface: Wireless environmental information sensing, feature, semantics, and knowledge,'' \emph{IEEE Commun. Mag.}, vol.~63, no.~8, pp. 56--62, Aug. 2025.

\bibitem{ITU2022FutureIMT2030}
{International Telecommunication Union}, ``Future technology trends of terrestrial international mobile telecommunications systems towards 2030 and beyond,'' ITU, Tech. Rep., 2022.

\bibitem{Liu2022JSACISACSurvey}
F.~Liu \emph{et~al.}, ``Integrated sensing and communications: Toward dual-functional wireless networks for {6G} and beyond,'' \emph{IEEE J. Sel. Areas Commun.}, vol.~40, no.~6, pp. 1728--1767, Mar. 2022.

\bibitem{10077117}
J.~Yang \emph{et~al.}, ``Multi-domain cooperative {SLAM}: The enabler for integrated sensing and communications,'' \emph{IEEE Wireless Commun.}, vol.~30, no.~1, pp. 40--49, Mar. 2023.

\bibitem{Leitinger2019BP_SLAM}
E.~Leitinger \emph{et~al.}, ``A belief propagation algorithm for multipath-based {SLAM},'' \emph{IEEE Trans. Wireless Commun.}, vol.~18, no.~12, pp. 5613--5629, Dec. 2019.

\bibitem{10962304}
S.~Zhai \emph{et~al.}, ``Multipath-based {SLAM} exploiting extended object estimation and classification,'' \emph{IEEE Trans. Wireless Commun.}, vol.~24, no.~8, pp. 7029--7045, Aug. 2025.

\bibitem{Sun2025LandToShipTWC}
S.~Sun \emph{et~al.}, ``Modeling and analysis of land-to-ship maritime wireless channels at 5.8 {GHz},'' \emph{IEEE Trans. Wireless Commun.}, vol.~25, pp. 10\,051--10\,065, 2026.

\bibitem{11352781}
X.~Cheng \emph{et~al.}, ``{APEG}: Adaptive physical layer authentication with channel extrapolation and generative {AI},'' \emph{IEEE Trans. Inf. Forensics Secur.}, vol.~21, pp. 1257--1272, 2026.

\bibitem{11153056}
Z.~Fang \emph{et~al.}, ``Environment reconstruction in terahertz monostatic sensing: Joint millimeter-level geometry mapping and material identification,'' \emph{IEEE J. Sel. Topics Electromagn., Antennas Propag.}, early access, 2025, doi:{\href{http://dx.doi.org/10.1109/JSTEAP.2025.3605128}{10.1109/JSTEAP.2025.3605128}}.

\bibitem{Chang2025WCLTHzEnvRecon}
Z.~Chang \emph{et~al.}, ``Environment reconstruction with multi-targets reflectors-merged sensing method based on {THz} single-sided channel characteristics,'' \emph{IEEE Wireless Commun. Lett.}, vol.~14, no.~5, pp. 1471--1475, May 2025.

\bibitem{Mou2023mmWave3DSLAMarXiv}
\BIBentryALTinterwordspacing
Z.~Mou and F.~Gao, ``Millimeter wave wireless communication assisted three-dimensional simultaneous localization and mapping,'' 2023. [Online]. Available: \url{https://arxiv.org/abs/2303.02617}
\BIBentrySTDinterwordspacing

\bibitem{Mahler2014Advances}
R.~P.~S. Mahler, \emph{Advances in Statistical Multisource-Multitarget Information Fusion}.\hskip 1em plus 0.5em minus 0.4em\relax Norwood, MA, USA: Artech House, 2014.

\bibitem{Kim2020mmWavePHD}
H.~Kim \emph{et~al.}, ``{5G} mmwave cooperative positioning and mapping using multimodel {PHD} filter and map fusion,'' \emph{IEEE Trans. Wireless Commun.}, vol.~19, no.~6, pp. 3782--3795, Jun. 2020.

\bibitem{Du2024IoTJGeneralSLAM}
T.~Du \emph{et~al.}, ``General simultaneous localization and mapping scheme for {mmWave} communication systems,'' \emph{IEEE Internet Things J.}, vol.~11, no.~12, pp. 22\,521--22\,536, 2024.

\bibitem{Kschischang2001FactorGraph}
F.~R. Kschischang \emph{et~al.}, ``Factor graphs and the sum-product algorithm,'' \emph{IEEE Trans. Inf. Theory}, vol.~47, no.~2, pp. 498--519, Feb. 2001.

\bibitem{10615625}
X.~Li \emph{et~al.}, ``A belief propagation algorithm for multipath-based {SLAM} with multiple map features: A {mmWave} {MIMO} application,'' in \emph{2024 IEEE International Conference on Communications Workshops (ICC Workshops)}, 2024, pp. 269--275.

\bibitem{Mendrzik2019SituationalAwareness}
R.~Mendrzik \emph{et~al.}, ``Enabling situational awareness in millimeter wave massive {MIMO} systems,'' \emph{IEEE J. Sel. Topics Signal Process.}, vol.~13, no.~5, pp. 1196--1211, Sep. 2019.

\bibitem{Leitinger2023DataFusionSLAM}
E.~Leitinger \emph{et~al.}, ``Data fusion for multipath-based {SLAM}: Combining information from multiple propagation paths,'' \emph{IEEE Trans. Signal Process.}, vol.~71, pp. 4011--4028, Sep. 2023.

\bibitem{Yang2022JSACHybridActivePassive}
J.~Yang \emph{et~al.}, ``Hybrid active and passive sensing for {SLAM} in wireless communication systems,'' \emph{IEEE J. Sel. Areas Commun.}, vol.~40, no.~7, pp. 2146--2163, Jul. 2022.

\bibitem{Hu2024RadioSLAM}
B.~Hu \emph{et~al.}, ``Multipath identification, user localization, and environment mapping in radio {SLAM},'' \emph{IEEE Trans. Commun.}, vol.~72, no.~10, pp. 6457--6473, Oct. 2024.

\bibitem{Sun2025RayleighDistance}
S.~Sun \emph{et~al.}, ``How to differentiate between near field and far field: Revisiting the {Rayleigh} distance,'' \emph{IEEE Commun. Mag.}, vol.~63, no.~1, pp. 22--28, Jan. 2025.

\bibitem{9898900}
Z.~Gao \emph{et~al.}, ``Integrated sensing and communication with {mmWave} massive {MIMO}: A compressed sampling perspective,'' \emph{IEEE Trans. Wireless Commun.}, vol.~22, no.~3, pp. 1745--1762, Sep. 2023.

\bibitem{Ge2020_5GSLAM_Sensors}
Y.~Ge \emph{et~al.}, ``{5G SLAM} using the clustering and assignment approach with diffuse multipath,'' \emph{Sensors}, vol.~20, no.~16, p. 4656, Aug. 2020.

\bibitem{Kim2022PMBM_VehTech}
H.~Kim \emph{et~al.}, ``{PMBM}-based {SLAM} filters in {5G} {mmWave} vehicular networks,'' \emph{IEEE Trans. Veh. Technol.}, vol.~71, no.~8, pp. 8646--8661, Aug. 2022.

\bibitem{Wielandner2023NonIdealSurfaces}
L.~Wielandner \emph{et~al.}, ``Multipath-based {SLAM} for non-ideal reflective surfaces exploiting multiple-measurement data association,'' \emph{J. Adv. Inf. Fusion}, vol.~18, no.~2, pp. 59--77, Dec. 2023.

\bibitem{Wielandner2024MIMO_NonIdealSurfaces}
------, ``{MIMO} multipath-based {SLAM} for non-ideal reflective surfaces,'' in \emph{Proceedings of the 27th International Conference on Information Fusion (FUSION)}, Venice, Italy, Jul. 2024, pp. 1--8.

\bibitem{Zhang2024ISACChannelComMag}
J.~Zhang \emph{et~al.}, ``Integrated sensing and communication channel: Measurements, characteristics and modeling,'' \emph{IEEE Commun. Mag.}, vol.~62, no.~6, pp. 98--104, Jun. 2024.

\bibitem{Liu2024SharedClusterISAC_TVT}
Y.~Liu \emph{et~al.}, ``A shared cluster-based stochastic channel model for integrated sensing and communication systems,'' \emph{IEEE Trans. Veh. Technol.}, vol.~73, no.~5, pp. 6032--6044, May 2024.

\bibitem{esposti2007measurement}
V.~Degli-Esposti \emph{et~al.}, ``Measurement and modeling of scattering from buildings,'' \emph{IEEE Trans. Antennas Propag.}, vol.~55, no.~1, pp. 143--153, Jan. 2007.

\bibitem{ZhangDiffuseScatteringNPJ_toappear}
T.~Zhang \emph{et~al.}, ``Diffuse scattering measurements and mechanism analysis at 8, 12, and 28 {GHz} for typical building surfaces,'' to appear in \emph{npj Wireless Technology}.

\bibitem{xie2016overview}
H.~Xie \emph{et~al.}, ``An overview of low-rank channel estimation for massive {MIMO} systems,'' \emph{IEEE Access}, vol.~4, pp. 7313--7321, 2016.

\bibitem{yang2018channel}
L.~Yang \emph{et~al.}, ``Channel estimation for millimeter-wave {MIMO} communications with lens antenna arrays,'' \emph{IEEE Trans. Veh. Technol.}, vol.~67, no.~4, pp. 3239--3251, Apr. 2018.

\bibitem{9693225}
B.~Barneto \emph{et~al.}, ``Millimeter-wave mobile sensing and environment mapping: Models, algorithms and validation,'' \emph{IEEE Trans. Veh. Technol.}, vol.~71, no.~4, pp. 3900--3916, Jan. 2022.

\bibitem{10600143}
B.~Lin \emph{et~al.}, ``Environment reconstruction based on multi-user selection and multi-modal fusion in {ISAC},'' \emph{IEEE Trans. Wireless Commun.}, vol.~23, no.~10, pp. 15\,083--15\,095, Jul. 2024.

\bibitem{Pei2025TrackingConvertedRadar}
\BIBentryALTinterwordspacing
A.~J. Pei, ``Tracking methods for converted radar measurements,'' \emph{Johns Hopkins APL Tech. Dig.}, vol.~38, no.~1, pp. 1--12, 2025, article 2503437. [Online]. Available: \url{https://www.jhuapl.edu/sites/default/files/2025-09/38-01-Pei.pdf}
\BIBentrySTDinterwordspacing

\bibitem{9585528}
F.~Meyer and J.~L. Williams, ``Scalable detection and tracking of geometric extended objects,'' \emph{IEEE Trans. Signal Process.}, vol.~69, pp. 6283--6298, Oct. 2021.

\bibitem{kay1993fundamentals}
S.~M. Kay, \emph{Fundamentals of Statistical Signal Processing: Estimation Theory}, ser. Prentice-Hall Signal Processing Series.\hskip 1em plus 0.5em minus 0.4em\relax Upper Saddle River, NJ, USA: Prentice-Hall, 1993, vol.~1.

\bibitem{bar-shalom2011tracking}
Y.~Bar-Shalom \emph{et~al.}, \emph{Tracking and Data Fusion: A Handbook of Algorithms}.\hskip 1em plus 0.5em minus 0.4em\relax Storrs, CT, USA: Yaakov Bar-Shalom, 2011.

\bibitem{ITUR_P2040_4_2025}
\BIBentryALTinterwordspacing
{International Telecommunication Union, Radiocommunication Sector (ITU-R)}, ``Effects of building materials and structures on radio-wave propagation in the range of {1} {MHz} to {450} {GHz},'' Recommendation ITU-R P.2040-4, Sep. 2025. [Online]. Available: \url{https://www.itu.int/dms_pubrec/itu-r/rec/p/R-REC-P.2040-4-202509-I%21%21PDF-E.pdf}
\BIBentrySTDinterwordspacing

\bibitem{3gpp_ts_22_125_v18_1_0}
\BIBentryALTinterwordspacing
{3GPP}, ``{3GPP TS 22.125 (ETSI TS 122 125): 5G; Unmanned Aerial System (UAS) support in 3GPP},'' European Telecommunications Standards Institute (ETSI), 3GPP Technical Specification TS 22.125 / ETSI TS 122 125, May 2024, version 18.1.0, Release 18, (2024-05). See Table 7.3-1 for positioning performance requirements. [Online]. Available: \url{https://www.etsi.org/deliver/etsi_ts/122100_122199/122125/18.01.00_60/ts_122125v180100p.pdf}
\BIBentrySTDinterwordspacing

\bibitem{3gpp_ts_22_186_v19_0_0}
\BIBentryALTinterwordspacing
------, ``{3GPP TS 22.186}: Service requirements for enhanced {V2X} scenarios,'' {ETSI}, Technical Specification TS 122 186, Oct. 2025, version 19.0.0 (Release 19). [Online]. Available: \url{https://www.etsi.org/deliver/etsi_ts/122100_122199/122186/19.00.00_60/ts_122186v190000p.pdf}
\BIBentrySTDinterwordspacing

\bibitem{5gamericas_isac_2025}
\BIBentryALTinterwordspacing
{5G Americas}, ``{Transforming Industries with Integrated Sensing and Communications},'' 5G Americas, White Paper, Jun. 2025, see Table 3.1 for example use-case requirements (e.g., AGV detection/tracking with ~10 cm location accuracy). [Online]. Available: \url{https://5gamericas.org/wp-content/uploads/2025/06/Transforming-Industries-with-Integrated-Sensing-and-Communications.pdf}
\BIBentrySTDinterwordspacing

\bibitem{horridge2011using}
P.~R. Horridge and S.~Maskell, ``Using a probabilistic hypothesis density filter to confirm tracks in a multi-target environment,'' in \emph{Proceedings of INFORMATIK}, Berlin, Germany, Jul. 2011, pp. 1--12.

\end{thebibliography}

%\endgroup

\end{document}